%% file: main.tex
\documentclass[conference]{IEEEtran}
\IEEEoverridecommandlockouts
\usepackage{cite}
\usepackage{amsmath,amssymb,amsfonts}
\usepackage{algorithmic}
\usepackage{graphicx}
\usepackage{textcomp}
\usepackage{xcolor}

\def\BibTeX{{\rm B\kern-.05em{\sc i\kern-.025em b}\kern-.08em
    T\kern-.1667em\lower.7ex\hbox{E}\kern-.125emX}}
\input{macros.tex}

\begin{document}

%

\title{Automatic Unit Test Generation for Deep Learning Frameworks based on API Knowledge}

\author{
    \IEEEauthorblockN{Arunkaleeshwaran Narayanan\IEEEauthorrefmark{1}, Nima Shiri harzevili\IEEEauthorrefmark{1}, Junjie Wang\IEEEauthorrefmark{2}, Lin Shi\IEEEauthorrefmark{2}, Moshi Wei\IEEEauthorrefmark{1}, 
    Song Wang\IEEEauthorrefmark{1}}
    \IEEEauthorblockA{\IEEEauthorrefmark{1}York University; \IEEEauthorrefmark{2}Institute of Software, Chinese Academy of Sciences
     \\\{arun98,nshiri,moshiwei,wangsong\}@yorku.ca,junjie@iscas.ac.cn,shilin@iscas.ac.cn}
}


\maketitle

\begin{abstract}
\input{sec/abstract}
\end{abstract}


\input{sec/introduction}

\input{sec/motivation}
\input{sec/approach}
\input{sec/setup}
\input{sec/result}
\input{sec/discussion}
\input{sec/related}

\input{sec/conclusion}


\balance
\bibliographystyle{IEEEtran}
\bibliography{paper}

\end{document}

%% file: macros.tex
\usepackage{soul}
\usepackage{hyperref}
\usepackage{multirow}
\usepackage{fancybox}
\usepackage{enumitem}
\usepackage{graphicx}
\usepackage{balance}
\usepackage{color}
\usepackage{float}
\usepackage{xcolor}
\usepackage{array}
\usepackage{amsmath}
\usepackage{xspace}
\usepackage{url}
\usepackage{tikz}
\usepackage{caption}
\usepackage{pgfplots}
\usepackage{listings}
\usepackage{color}
\usepackage{graphicx}
\definecolor{dkgreen}{rgb}{0,0.6,0}
\definecolor{gray}{rgb}{0.5,0.5,0.5}
\definecolor{mauve}{rgb}{0.58,0,0.82}
\pgfplotsset{compat=1.16}
\usepackage{pgf-pie}
\usetikzlibrary{pgfplots.statistics,calc}
\usepackage{algorithm}
\usepackage{algorithmic}
\usepackage{subcaption}
\usepackage{listings}

\usepackage{tcolorbox}

\makeatletter
\newcommand{\mybox}[1]{%
	\setbox0=\hbox{#1}%
	\setlength{\@tempdima}{\dimexpr\wd0+13pt}%
	\begin{tcolorbox}[boxrule=0.5pt, colback=white, arc=4pt,
		left=6pt,right=6pt,top=6pt,bottom=6pt,boxsep=0pt]
		#1
	\end{tcolorbox}
}

\definecolor{codegreen}{rgb}{0,0.6,0}
\definecolor{codegray}{rgb}{0.5,0.5,0.5}
\definecolor{codepurple}{rgb}{0.58,0,0.82}
\definecolor{backcolour}{rgb}{0.95,0.95,0.92}

\lstdefinestyle{mystyle}{
  language=Python,
  aboveskip=3mm,
  showstringspaces=false,
  columns=flexible,
  numbers=none,
  backgroundcolor=\color{backcolour},
  commentstyle=\color{codegreen},
 keywordstyle=\color{magenta},
    numberstyle=\tiny\color{codegray},
    stringstyle=\color{codepurple},
    basicstyle=\small\ttfamily,
    breakatwhitespace=false,         
    breaklines=false,                 
    captionpos=b,                    
    keepspaces=false,                 
    numbersep=5pt,                  
    showspaces=false,                
    showstringspaces=false,
    showtabs=false,                  
    tabsize=2,
    escapeinside=``
}
\definecolor{songcolor}{RGB}{191,191,191}
\definecolor{nimacolor}{RGB}{0.13, 0.67, 0.8}
\definecolor{aruncolor}{RGB}{51,255,51}
\newcommand{\todoc}[2]{{\textcolor{#1}{\textbf{#2}}}}
\newcommand{\todored}[1]{\todoc{red}{\textbf{[[#1]]}}}

\newcommand{\song}[1]{\todored{Song: #1}}

 \newcommand{\arun}[1]{\todored{arun: #1}}

\newcommand{\tool}{\texttt{MUTester}}

%% file: sec/abstract.tex
{
{Many automatic unit test generation tools that can generate unit test cases with high coverage over a program have been proposed.  However, most of these tools are ineffective on deep learning (DL) frameworks due to the fact that many of deep learning APIs expect inputs that follow specific API knowledge.} 

To fill this gap, we propose {\tool} to generate unit test cases for APIs of deep learning frameworks by leveraging the API constraints mined from the corresponding API documentation and the API usage patterns mined from code fragments in Stack Overflow (SO).  
Particularly, we first propose a set of 18 rules for mining API constraints from the API documents. We then use the frequent itemset mining technique to mine the API usage patterns from a large corpus of machine learning API related code fragments collected from SO. 
Finally, we use the above two types of API knowledge to guide the test generation of existing test generators for deep learning frameworks.} 
To evaluate the performance of {\tool}, we first collect 1,971 APIs from four widely-used deep learning frameworks (i.e., Scikit-learn, PyTorch, TensorFlow, and CNTK) and for each API, we further extract its API knowledge, i.e., API constraints and API usage. Given an API, {\tool} combines its API knowledge with existing test generators (e.g., search-based test generator PyEvosuite and random test generator PyRandoop) to generate test cases to test the API. 
Results of our experiment show that {\tool} can significantly improve the corresponding test generation methods and the improvement in code coverage is 15.7\% to 27.0\% on average. 
In addition, it can help reduce around 19.0\% of invalid tests generated by the existing test generators. 
{Our user study with 16 developers further demonstrates the practicality of {\tool} in generating test cases for deep learning frameworks.}


%% file: sec/introduction.tex
\section{Introduction}
\label{sec:intro}

To assure software quality,
many automated approaches that explore the input space and generate effective unit test cases for given APIs have been proposed in recent years, e.g., Evosuite~\cite{fraser2011evosuite}, i.e., a typical search-based test generation technique and Randoop~\cite{pacheco2007randoop,pacheco2007feedback}, i.e., a feedback-directed random test generation approach.
These tools have been showed are effective at generating unit test cases with high code coverage and can be used to help developers write high-quality unit test cases for APIs of traditional software systems~\cite{fraser2011evosuite}.  
However, recent studies show that the performance of these test case generation tools on DL frameworks are limited due to the fact that many of DL APIs expect inputs that follow DL-specific API knowledge, i.e., input constraints and API usages, 
which cannot be provided by most of the existing test generation approaches~\cite{song2021how,zhang2020machine}.
\input{figure/APIknowledgeExample}
\input{figure/testCaseExample}

Specifically, most of the existing test generators mainly target maximizing the code coverage under different strategies when synthesizing method call sequences or inputs for testing an API without considering {its constraints}. 
However, generating valid and effective unit test cases for DL APIs requires two types of API knowledge: 1) the input constraints, i.e., the data structures of the input parameters such as arrays, lists, and tuples. 
For example, Figure~\ref{fig1:label1} shows that \texttt{fit()} method of KMeans in Scikit-Learn framework has three parameters, and each of them has different constraints, e.g., the first parameter \texttt{X} must be a two dimension array of numerical values, otherwise the function call to this method will crash; 2) API usage, i.e., the context of an API determined by the nature of the process of DL tasks. For example, Figure~\ref{fig1:label2} shows two possible API usage patterns of the method \texttt{fit()}, in both example we can see that before calling method \texttt{predict(X)} to predict the closest cluster each sample in \texttt{X} belongs to, one needs to call method \texttt{fit()} to compute k-means clustering. Ignoring the API usages results in generating many invalid test cases.

Most traditional test case generation tools often suffer from lack of the API knowledge, due to which they tend to generate test cases with invalid input values or incorrect method call sequences. 
For example, Figure~\ref{fig:2a} shows a test case developed by a search-based test case generation approach for \texttt{sklearn.cluster.KMeans}. 
The variables generated are tend to be in no accordance with the constraints of \texttt{sklearn.cluster.KMeans()}, e.g., the method requires its first parameter to be a numerical value and it is also a required parameter, but the generated test case passes a \texttt{None}, which makes the generated test case invalid.  
In addition, before calling \texttt{predict(X)}, one needs to call method \texttt{fit()} to compute k-means clustering, otherwise the generated API scenario is incorrect.  
This nature of lack of API knowledge leads existing test case generators produce many invalid test cases, since Python is a dynamic programming language, such invalid tests could pass without throwing any error, which potentially consumes the testing time yet without improving the coverage.

{To address these limitations of existing test case generators, this paper proposes {\tool} to generate test cases for APIs of DL frameworks by leveraging the API constraints mined from API documents and API usage patterns mined from code fragments in Stack Overflow.}  
Figure~\ref{fig:2b} shows a test case generated by {\tool}. 
We can see that the API is fed with valid inputs, generated with the guidance of the mined API constraints of \texttt{KMeans}. 
In addition, with the mined API usage patterns, the generated test case also has valid method calls to other APIs, 
which makes the test case more practical, thus achieving better code coverage. 




{We evaluate {\tool}, in guiding two commonly-used automatic test case generation tools, i.e., the search-based test generation tool \texttt{Evosuite} and the random test generation tool \texttt{Randoop}.
We utilize the Python version of these tools (i.e., \textit{PyEvosuite} and \textit{PyRandoop}) implemented by the authors of Evosuite in~\cite{lukasczyk2020automated,journals/corr/abs-2202-05218}.}
To evaluate the performance of {\tool}, we first collect 1,971 APIs from four widely-used DL frameworks (i.e.,  Scikit-learn~\cite{sklearn_api}, 
PyTorch~\cite{NEURIPS2019_9015}, TensorFlow~\cite{abadi2016tensorflow}, and CNTK~\cite{seide2016cntk}) and for each API, we further extract its API knowledge, i.e., API constraints from API documents and API usage patterns from SO code examples. Given an API, {\tool} combines its API knowledge and the existing test case generation method (e.g., search-based test generation or random test generation) to generate test cases for the API. 
Results of our experiment on the four DL frameworks show that {\tool} can significantly improve existing test case generation methods, {i.e., increasing 20.5\% code coverage and decreasing 19.0\% invalid tests generation. } 
{Our user study with 16 developers further demonstrates that {\tool} can generate more readable and useful unit test cases for DL frameworks than the baselines, which show the practice value of {\tool}.} 
This paper makes the following contributions:

\begin{itemize}

\item We proposes {\tool} to generate test cases for
APIs of DL frameworks by leveraging the API knowledge mined from API documents and code examples in Stack Overflow.

\item
We develop a set of 18 linguistic rules for mining the API constraints from the API documentation and we leverage frequent itemset mining technique to mine API usage patterns from code fragments in Stack Overflow. 

\item Our evaluation on four typical DL frameworks {shows the effectiveness of {\tool} in improving code coverage and reducing invalid test cases. 
The user study also demonstrates the usefulness and practicality of our tool.}

\item We release the source code of {\tool} and the dataset of our
experiments to help other researchers replicate and extend
our study\footnote{https://zenodo.org/record/7987893}.


\end{itemize}

The rest of this paper is organized as follows. 
Section~\ref{sec:motivation} presents
the background of this work.  
Section~\ref{sec:approach} describes the methodology of our approach. 
Section~\ref{sec:experiment} and Section~\ref{sec:result} respectively show the experimental setup and the evaluation results. 
Section~\ref{sec:discussion} discusses the threats to validity of this work. 
Section~\ref{sec:related} presents the related studies. 
Section~\ref{sec:conclusion} concludes this paper.


%% file: figure/APIknowledgeExample.tex
\begin{figure}[H]
	\centering
	\begin{subfigure}{.48\columnwidth}
		\centering
		\includegraphics[width=\textwidth]{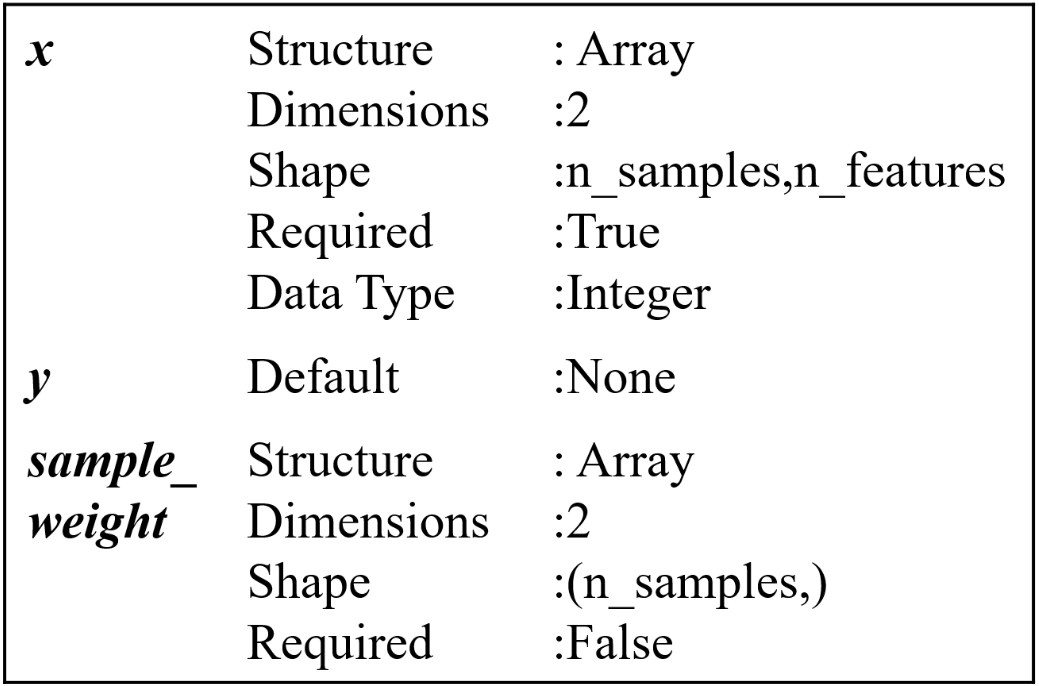}
		\caption{API constraints}
		\label{fig1:label1}
	\end{subfigure}%
	\hfill
	\begin{subfigure}{.48\columnwidth}
		\centering
		\includegraphics[width=\textwidth]{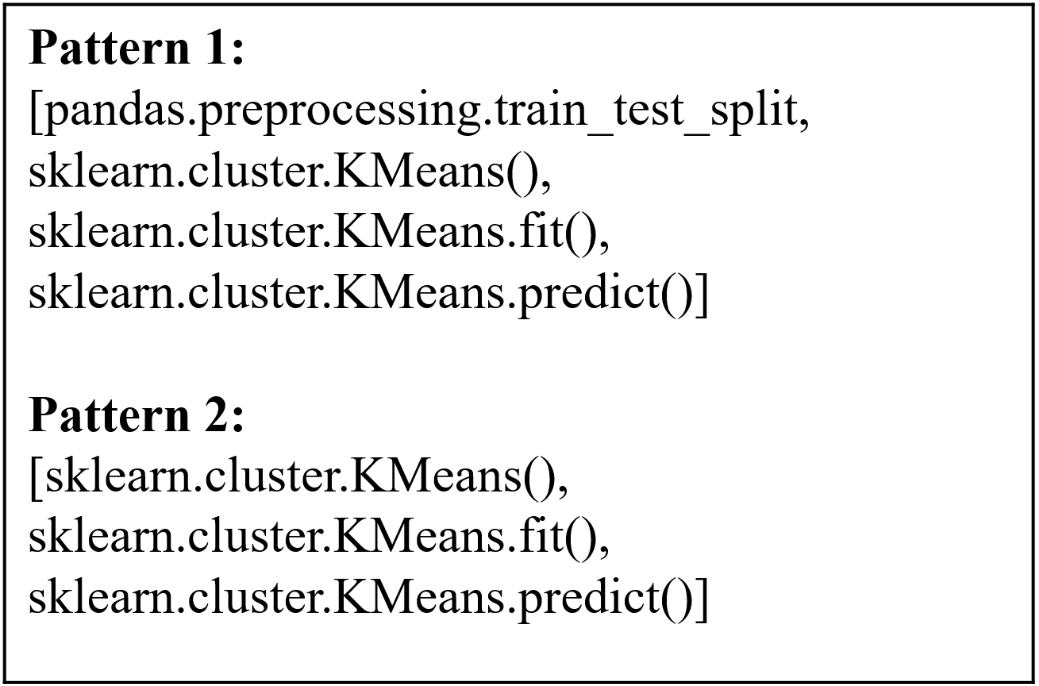}
		\caption{API usage patterns}
		\label{fig1:label2}
	\end{subfigure}
 \vspace{-0.1in}
	\caption{\small API Knowledge of method \texttt{fit(X, y=None, sample\_weight=None)} in \texttt{sklearn.cluster.KMeans}.}
	\label{fig:Apiknowledge}
\end{figure}

%% file: figure/testCaseExample.tex
\begin{figure*}[t!]
	\centering
	\begin{minipage}{0.5\textwidth}
		\centering
		\begin{lstlisting}[language=Python]
	import k_means as module_0
    def test_case():
        var0 = None
        var1 = [var0]
        var2 = 'm!"%SL2@D'
        var3 = None
        var4 = '6qRGnw"e='
        var5 = "k=ft\n<'?~GS\x0c|iD^"
        var6 = 'cYx\x0b< _RI5R'
        var7 = '\x0bD'
        var8 = var2, var5, var6, var7
        var9 = module0.KMeans(var3,n_init=var3,
                max_iter=var0,n_jobs=var4,algorithm=var8)
        var10 = module0.predict(var3)
        assert var9 is not None
		\end{lstlisting}
		\subcaption[second caption.]{Test case generated by \textbf{PyEvosuite}.}\label{fig:2a}
	\end{minipage}%
	\begin{minipage}{0.5\textwidth}
		\centering
		\begin{lstlisting}[language=Python]
	import k_means as module_0
	import _split as module_1

    def test_case():
        int_0 = 19
        var_0 = module_0.KMeans(int_0)
        float_0 = 1.6484
        int_2 = 69672
        float_1 = 83.14
        int_3 = 5568
        list_0 = [[float_0,int_2],[float_1,int_3]]
        var_1 = module_1.train_test_split(list_0)
        var_2 = var_0.fit(list_0)
        var_3 = var_0.predict(list_0)
        assert var_3 is not None
		\end{lstlisting}
		\subcaption[second caption.]{Test case generated by {\tool}.}\label{fig:2b}
	\end{minipage}%
\vspace{-0.1in}
	\caption{Test cases generated by \textbf{PyEvosuite} and {\tool}.
	} 
	\label{fig:testcaseexample}
\end{figure*}

%% file: sec/motivation.tex
\section{Background on Automatic Unit Test Generation}
\label{sec:motivation}
Automated unit test generators share a common process to generate test cases for a given API, i.e., synthesizing method call sequences that involves this API (i.e., synthesizing usage scenario) and generating inputs for parameters of each method in the call sequences(i.e., synthesizing inputs) ~\cite{fraser2012whole,derakhshanfar2020search,olsthoorn2020generating,mcminn2012search,shahbaz2015automatic,arcuri2021evomaster,atlidakis2019restler}.  
Although there exist many techniques for automatic test generation, this paper focuses on two types of widely-adopted and scalable approaches for test case generation, i.e., random test generation and search-based test generation. 

\subsubsection{Random Test Generation}
Random testing is a basic and scalable approach
for test generation~\cite{arcuri2011random}, which generates test cases by creating the invocation of functions with random inputs. Guided random testing is a refined approach that starts with random input data, then uses extra
knowledge to guide input data generation. One typical example of random testing is feedback-directed random testing, i.e., Randoop~\cite{pacheco2007randoop,pacheco2007feedback}, 
which improves the random test generation by analysing the feedback collected from the previously generated test cases to avoid illegal input data. It starts by generating a random input data, then uses the feedback knowledge to develop input generation. Each of the generated test cases is executed immediately, to derive feedback and generate new ones. 

\subsubsection{Search-based Test Generation}
\label{sec:searchtech}

 Search-based test generation employs a genetic algorithm, that uses evolutionary operators like crossover, mutation, and selection to iteratively improve the candidate solutions for optimization of the fitness function. Fitness function in the search-based unit test generation often comprise code coverage of the generated test cases. 
 Evosuite~\cite{fraser2011evosuite} is one of the typical search-based unit test generation tool. 

Note that, most of the existing test generation approaches, i.e., Evosuite~\cite{fraser2011evosuite} and Randoop~\cite{pacheco2007randoop,pacheco2007feedback}, are designed for object-oriented programming language such as Java and cannot be directly applied to the studied Python-based deep learning frameworks. 
In this work, we use the Python version search-based test generator (denoted as \textbf{PyEvosuite}) and random test generator (denoted as \textbf{PyRandoop}) implemented   in~\cite{lukasczyk2020automated} as our experiment subjects.   





%% file: sec/approach.tex
\section{Approach}
\label{sec:approach}

\input{figure/overview}


Figure~\ref{fig:overview} shows the pipeline of {\tool}, 
which consists of three steps, i.e., mining API constraints from API documents (Section~\ref{sec:apiCons}), mining API usage patterns from code fragments in Stack Overflow (Section~\ref{sec:mineAPIusage}), and leveraging the API constraints and API usage patterns to generate test cases (Section~\ref{sec:testCaseGene}).


\subsection{API Constraints Mining from API Documentation}
\label{sec:apiCons}

\input{table/Patter_t}

To help the end-users {take full advantage of the APIs, the API documentation often provide publicly available information about the usage of each API \cite{9054844} with the input formats and example code snippets, and etc. } 
This information can be helpful for generating test cases with API input constraints {and facilitating the generation of valid inputs.}  
However, manually analysing API documents and extracting constraints of APIs is time and effort consuming, 
e.g., Scikit-Learn has 1,210 APIs with each having at least one parameter and in some cases like API \texttt{sklearn.cluster.KMeans} has 22 parameters. 
To tackle this challenge, we propose a semi-automated approach to collect API constraints, 
{in which we first manually design two static analyzers and 18 linguistic rules} regarding API constraints in the data structure, shape/size/dimension of data structure, data type, value/default value, and optionality of parameters, {then we apply these rules for automatic extraction of API constraints.}
In this paper, the API constraints mainly refer to the input constraints of an API, i.e., the data structures and properties of these data structures of inputs. 

Specifically, 
given an API, {we collect its API constraints from the following three sources. } 
\textbf{1) Signature definition}, 
i.e., the definition line for a class or a function, which shows the name of the class/function, and all the required and optional parameters. The signature definition also includes the default values for the optional parameters. For example, Figure~\ref{fig:headerLine} shows the signature definition of API  \texttt{fit} from \texttt{sklearn.cluster.KMeans}. 
\textbf{2) Parametric page}, i.e., the natural language sentences that describe the usage of a parameter with information about its data type, data range, fixed values, size and shape details in cases of arrays, matrix, list, and others. These sentences are normally given for each parameter of an API. For example, Figure~\ref{fig:parametricPage} shows the parametric page of  API \texttt{fit} from \texttt{sklearn.cluster.KMeans}, from which we can infer that the parameter \texttt{X} is an array.   
\textbf{3) Example code}, i.e., the examples provided by a DL framework that contain the usage of some APIs with concrete inputs. 
For each example code, we extract the concrete inputs and input types of data structures indicated in the parameter pages of the involved APIs. 
  For example, Figure~\ref{figure:codeExamp} shows the example code of 
\texttt{sklearn.cluster.KMeans}.
To extract these documents data from the API documents, we designed a web scrapping tool using a python based HTML parser~\cite{richardson2007beautiful}.

The details of mining different types of API knowledge from these three sources are as follows.

\input{figure/headerline}
\input{figure/parameterLine}
\input{figure/codeExample}
                



\subsubsection{\textbf{Mining Signature Definition}} 
As the format of the most signature definitions is in a structured format, we developed a regular expression based string analyzer to mine and extract the following information from the signature definition, i.e., API name, required parameters (i.e., these parameters are the ones in the signature definition without default values and an API cannot be used without passing these required parameters), optional parameters (i.e., the parameters in signature definition with default values) with their default values. 

For example, given the signature definition of method \texttt{fit} showed in Figure~\ref{fig:headerLine}, our tool first identifies the API name is `fit' and it has three parameters, i.e., `X', `y', and `sample\_weight'. In addition,  `X' is a required parameter as there is no default value for it in the header line. Both `y' and `sample\_weight' are optional and the default values are `None'.



\subsubsection{\textbf{Mining Parametric Page}}

Typically a Python API has one natural language described parametric sentence for each of its parameters in DL frameworks to help users. {We observe that, these natural language descriptions of parameters are of various formats and styles, which brings challenges for automatically extracting the parameter related information. 
Nevertheless, we also find that they would share similar linguistic patterns when mentioning the data type, data structure, etc, which motivated us in designing linguistic rules for the automatic extraction.} 

To understand and design linguistic rules from the parametric pages, 
we have conducted a manual analysis on documents from 100 randomly selected APIs from each of the four DL frameworks. In total 400 APIs were analyzed.  
 \input{table/constraintsFit}
For our analysis, we first extracted all the parametric pages from the API documents of the studied 400 APIs, then we applied a pre-processing to remove noise tokens, e.g., white spaces and special characters~\cite{chowdhary2020natural}. 
After that, the first three authors worked together to summarize possible regular expression-based linguistic patterns 
for each parameter and derive an initial list of rules for mining constraints for the parameter. 
Then all the authors discussed the categorization results and merged similar ones, as a result 18 rules are summarized. 
These parametric rules are listed out with their examples in Table~\ref{tab:ruless}. 
For example, rule No.1 in Table~\ref{tab:ruless}, a common pattern that appears in the parametric pages of the studied DL API documents, can be used to extract two types of API constraints from the matched example ``int default=0'', i.e., the data type and the default value of the parameter, ``int'' and ``0'' respectively.  

\subsubsection{\textbf{Mining Example Code}}
\label{sec:mineExample}
To mine the example code, we build a heuristic-based static analyser to extract all the concrete input values of involved DL APIs from the example code, which can help us to precisely infer proprieties of APIs' input constraints, e.g., the concrete shape of the input data structure and the type of elements required in the input data structure. 
Note that not all the APIs have an example code snippet maintained by DL frameworks, for these APIs that do not have corresponding example code snippets, we reuse the proprieties of inputs from other APIs that share the same parameter pages. 
Given the example code showed in Figure~\ref{figure:codeExamp}, from the parametric page, we only know that the parameter \textit{X} is an array, yet from this example code, we can further infer that it should be a 2-D integer array, which can facilitate inputs synthesizing with precise data information during test generation. 

Table~\ref{tab:constraintsExample} shows the constraints mined for method \texttt{fit} of \texttt{sklearn.} \texttt{cluster.KMeans}, in which we can find the constraints for each of its three parameters. For example, parameter \texttt{X} is a required parameter and requires a 2D integer array.
Once all the input constraints are mined, they will be used to generate test case for DL APIs in Section~\ref{sec:testCaseGene}.

\subsection{API Usage Patterns Mining from SO}
\label{sec:mineAPIusage}
{The common usage patterns of APIs can provide important clues for achieving a programming task and facilitating generating valid } test cases~\cite{shamshiri2015automatically,song2021how}. 
Most of the existing test case generation techniques~\cite{arcuri2011random,fraser2011evosuite} randomly generate API call sequences to synthesize test cases, yet machine learning APIs are often used with specific order for different machine learning tasks. For example, given the two method \texttt{fit()} and \texttt{predict(X)} of\texttt{sklearn.cluster.KMeans} in Scikit-Learn framework, before calling \texttt{predict(X)} method to predict the closest cluster each sample in \texttt{X} belongs to, one needs to call method \texttt{fit()} to compute k-means clustering. Such API usage patterns can be used to help test case generation with valid API call sequences {and improve the coverage of the generated tests.}

To generate valid API call sequences for DL  frameworks, following existing studies~\cite{zhang2018code,uddin2020mining}, we use a frequent itemset mining based technique to mine API usage patterns from Stack Overflow code examples, 
which consists of two steps, i.e., API usage collection (Section~\ref{sec:SOCollection}) and API pattern mining (Section~\ref{sec:frequentItemset}). 
Note that, following existing
work~\cite{huang2018api,uddin2020mining}, we only collect code from the accepted answers of a question in SO for removing incorrect API calls. 


\subsubsection{\textbf{API Usage Collection}}
\label{sec:SOCollection}

We first select questions with the name of the four DL frameworks as Stack Overflow question tags to retrieve questions and only questions having accepted answers remain for removing incorrect API calls.  
We develop a heuristic-based APIs extraction method from the code fragments of accepted answers. Note that, only APIs from these studied DL frameworks are kept. 
If a question contains multiple APIs, we concatenate all APIs into
a list of APIs. 
As a result, we collected a total of 89,969 question-APIs pairs, which involve 1,971 unique APIs from the four studied DL frameworks. 
{After this, we obtain the API usage dataset which contains a set of API call sequences with each element being a list of APIs for achieving a specific programming task. } 

\subsubsection{\textbf{Mining API Usage Patterns}}
\label{sec:frequentItemset}
With the collected DL API usage dataset, we further mine the {API usage patterns reflecting the co-occurrence relations of the APIs.
We utilize the association rule mining technique on the API call sequences collected from the answers of ML programming tasks. }
Specifically, following existing study~\cite{zhong2009mapo}, we use the Apriori Algorithm~\cite{agrawal1994fast} to create a set of API association rules (i.e., API usage patterns) from the DL API usage dataset.  
We utilize the API call sequence from the answer of each SO post as input to the Apriori algorithm for rule mining. 
Apriori generates a pattern hypothesis by randomly selecting two or more APIs as the base itemset \textit{A} and then randomly selecting one or more APIs as the extension itemset \textit{B}. It further calculates the confidence and support of the pattern hypothesis
$\{A\Rightarrow B\}$ and compares it to the confidence and support threshold specified by the user. 
Apriori accepts a association rule if the \textit{confidence} and \textit{support} of the rule are higher than the threshold values specified~\cite{agrawal1994fast}. 
Specifically, \textit{support} indicates the frequency of an API association rule with respect to the entire dataset.  \textit{Confidence} indicates the
percentage of one or more extended API(s) (e.g., item set \textit{B}) found to be true given a base rule (e.g., item set \textit{A}). 

In our experiment, we empirically set \textit{support} to 2 (e.g., the minimum support value in Apriori) and set \textit{confidence} to 50\% to {potentially recall more API usage patterns.}

After that we rank the patterns of each API based on their confidence values. These patterns will be further used in {\tool} to help generate valid API call sequences. 
Note that, one can use small \textit{confidence} values, i.e., smaller than 50\%, to initialize the pattern generation process, while in our study, we find that a small \textit{confidence} value does not change test cases generated by {\tool} given a reasonable time budge, i.e., 5 minutes for each API (as the top patterns are the same), while running Apriori with a small \textit{confidence} value requires much more extra time. 
{After this step, we output a list of API usage patterns descending ranked by the confidence values for guiding the test generation. }



\subsection{API Knowledge Guided Test Generation}
\label{sec:testCaseGene}

After we extract the two types of API knowledge, i.e., API constraints (Section~\ref{sec:apiCons}) and API usage patterns (Section~\ref{sec:mineAPIusage}), {\tool} further leverages them to guide the test generation of existing approaches for DL frameworks.
Given a module from a DL framework, existing test generators share a common workflow to generate tests which contains two major steps, i.e., method call sequence synthesizing, input synthesizing for these invoked methods. In this paper, we use the mined API knowledge to guide these two steps with a target of generating more useful and valid tests given the same resource budget. Specifically, we leverage API usage patterns to narrow the search space for call sequence generation (see Section~\ref{sec:inteAPIUsage}) and use API constraints to help solve the type information for input generation (see Section~\ref{sec:inteAPIConstraints}).

\subsubsection{\textbf{API Usage Patterns Guided Call Sequence Generation}}
\label{sec:inteAPIUsage}
Given a module from a DL framework, existing test generators (i.e., \textbf{PyEvosuite} and \textbf{PyRandoop}) first parse the module and extract information about available methods in the module 
and its imports. The method call sequence construction often starts from creating the constructor of the module, during the execution, all the involved objects, methods, data, statements, etc., will be stored for further use. After that under a specific criterion (e.g., \textbf{PyEvosuite} targets at maximizing code coverage when choosing the next methods), different methods will be added into the call sequence. Such process end when stopping conditions are triggered (e.g., time limitation, code coverage satisfied, maximum statements executed). 

Current test generators do not consider the real-world API usage scenarios when constructing the method call sequence, thus can generate tests with incorrect API usages as shown in Figure~\ref{fig:2a}, i.e., before calling \texttt{predict(X)}, one needs to call method \texttt{fit()}, when using a K-Means algorithm provided by \texttt{sklearn.cluster.KMeans}. 
To solve this issue, {\tool} leverages API usage patterns to guide the method call construction. Specifically, for a under expanding method call sequence $M$, {\tool} will first iterate the available candidate method set and check whether a pattern can be matched based on $M$, if not, move to the longest sequential sub-sequence methods of $M$ {to check the usage patterns for the remaining methods} and recursively. If multiple candidate methods can be matched, {\tool} selects the one that has larger confidence and appends it to $M$. If no candidate method can be matched, {\tool} uses the criterion of the adopted test generator to select the next method to be appended to $M$. Details are shown in Algorithm~\ref{alg:algorithm1}.

\begin{algorithm}[t]
	\caption{API usage guided call sequence construction}
	\label{alg:algorithm1}
	\hrule
	\begin{algorithmic}[1]
		\vspace{.1cm}
		\REQUIRE ~\\
		Call sequence under expanding $M$ \\
		The available candidate method set $C$\\
		API pattern set $P$\\
		\ENSURE ~\\
		Next method to be appended to $M$
		\STATE initialize a priority method set $M_{priority}$
		\STATE initialize a backup method set $M_{backup}$
		\FOR{each candidate method $c$ in $C$}
		    \WHILE{$M$.size() >1}
			    \IF{\{$M$ $=>$ $c$\} is in $P$} 
			        \STATE put $c$ into $M_{priority}$
			        \STATE BREAK
                \ENDIF
                \STATE $M = M.subList$(1;)
            \ENDWHILE
            \IF {$c$ is not in $M_{priority}$}
                \STATE put $c$ into $M_{backup}$
            \ENDIF
		\ENDFOR
		\STATE sort $M_{priority}$ by confidence of patterns in $P$
        \STATE sort $M_{backup}$ by {the criterion in test generator}
        \IF{$M_{priority}$ is not empty}
            \RETURN top element in $M_{priority}$
        \ELSE 
        \RETURN top element in $M_{backup}$ 
		\ENDIF
	\end{algorithmic}
	\hrule
\end{algorithm}

\subsubsection{\textbf{API Constraints Guided Input Generation}}
\label{sec:inteAPIConstraints}
When a method is determined to be added into the method call sequence, a test generator needs  to generate inputs for the method. Current generators such as \textbf{PyEvosuite} and \textbf{PyRandoop} mainly leverage the objects, data, and variables collected from last execution to initialize the parameters of the method and randomly guess the type information and inputs for parameters that are not available from the last executions, while DL APIs often require more complex and framework  specific structures like numpy.array, tuple, Date\& time, tensor, etc., which often cannot be obtained from existing executions. 

{\tool} solves this issue by leveraging the API constraints mined from API documents. Specifically, given a parameter $p$ from a method $m$, {\tool} first obtains its basic type information regarding the data structure, the shape/size of data structure, data type, default value, etc., mined from $m$'s header line, parameter page, and example as illustrated in Section \ref{sec:apiCons}.
{\tool} then searches the last executions to see whether there exists matched inputs, if cannot find matches, {\tool} further randomly generates inputs by following the constraints. 
Note that, as shown in Section~\ref{sec:mineExample}, some of the concrete data type information of a specific data structure used in a method can only be inferred if the method appears in an example code, however not all APIs have a corresponding code example,  following the existing test generators (e.g., \textbf{PyEvosuite} and \textbf{PyRandoop}), we randomly guess a primitive data type for them.

%% file: figure/overview.tex
\begin{figure}[t!]
    \centering
      \includegraphics[width=0.85\columnwidth]{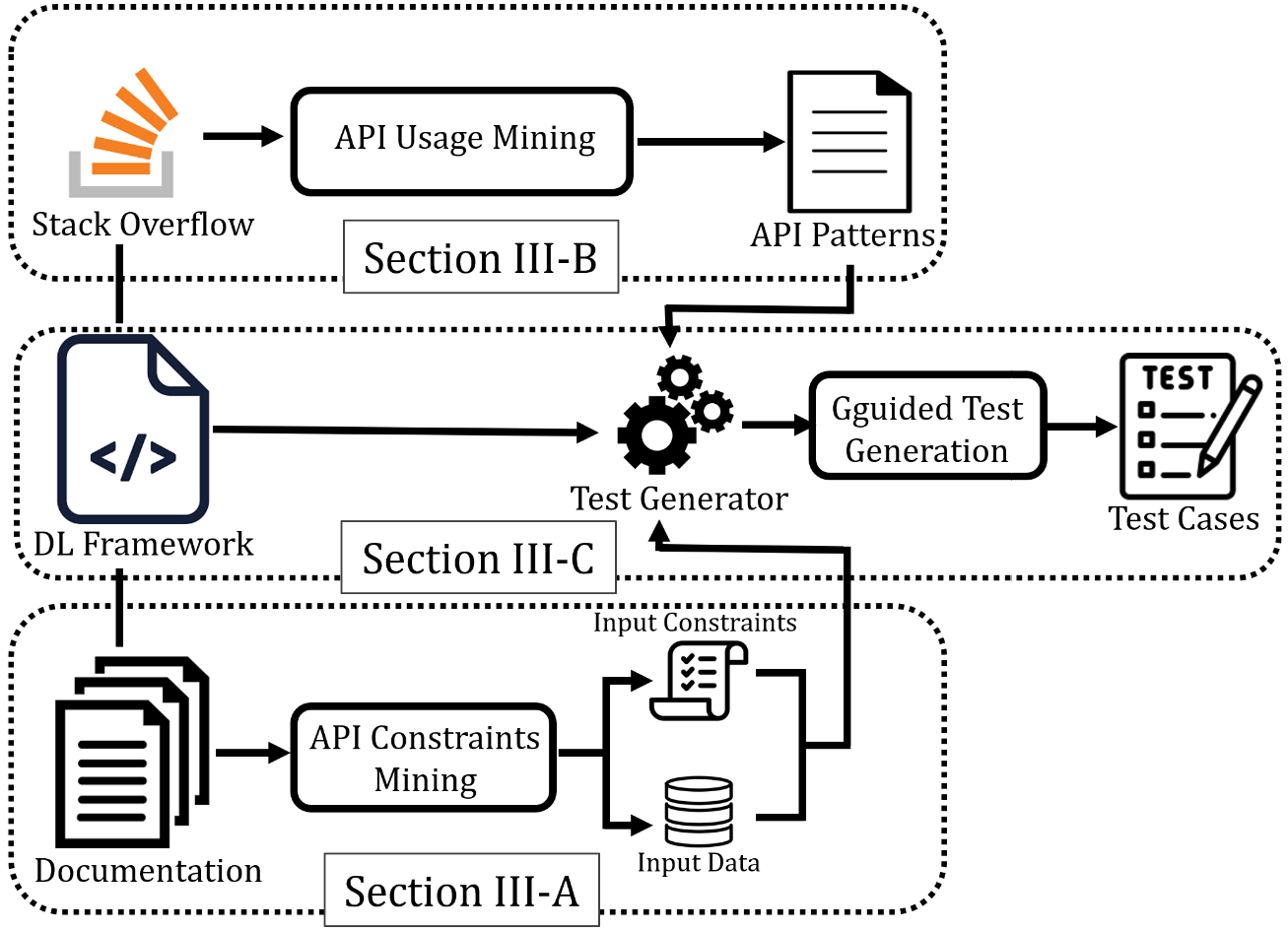}
      
    \caption{Overview of {\tool} 
    }
   \label{fig:overview}
\end{figure}

%% file: table/Patter_t.tex
\begin{table*}[t!]
\caption{Linguistic rules for mining constraints from parametric pages of APIs.}
\label{tab:ruless}
\begin{tabular}{llll}
\hline
No & Constraint Type & Linguistic Rules & Examples                   \\ \hline
1 &  Data type, Default Value  & \textless{}D\_type\textgreater \, default=\textless{}value\textgreater{} & int default=0                       \\
2  &  Data type & \textless{}D\_type\textgreater \, or \textless{}D\_type\textgreater{}    & int or float                        \\
3  &  Structure, Shape, Default Value&
  \textless{}Structure\textgreater \, of shape \textless{}(shape)\textgreater{}, default=\textless{}value\textgreater{} &
  array-like of shape (n\_samples,n\_features), default=None \\
4 & Structure, Shape, Default Value &
  \textless{}Structure\_Enum\textgreater \, of shape \textless{}(shape)\textgreater{}, default=\textless{}value\textgreater{} &
  \{array-like, sparse matrix\} \, of shape (n\_samples, n\_features), default=None \\
5 & Structure, Shape&
  \textless{}Structure\textgreater \, of shape \textless{}(shape)\textgreater \, or \textless{}(shape)\textgreater{} &
  array-like of shape (n\_samples,n\_features) \, or (n\_samples,) \\
6  & Structure, Data Type& \textless{}D\_type\textgreater{}/\textless{}structure\textgreater{}   & params: dict                        \\
7  & Structure & \{Structure\_Enum\}                                                   & \{list, tuple, set\}                \\
8  & Values, Default Value& \{Values\_Enum\} default=\textless{}value\textgreater{}               & \{`text', `diagram'\}, default=None \\
9  & Values & \{Values\_Enum\}                                                      & \{`text', `diagram'\}               \\
10 & Size, Structure &\textless{}Size\textgreater length of \{Structure\_Enum\}             & 2-length sequence (tuple, list, …)  \\
11 & Dimension, Structure &\textless{}Dimension\textgreater \, d \textless{}structure\textgreater{} & 2d Array                            \\
12 & Deafult Value, Values &
  \textless{}value\textgreater (def), \textless{}value\textgreater{}, .... or \textless{}value\textgreater{} &
  `backward' (default), `forward', or `nearest' \\
13 & Structure, Data Type &\textless{}Structure\textgreater \, of \textless{}D\_type\textgreater{}  & tuple of ints                       \\
14 & Values &\textless{}value\textgreater \, or \textless{}value\textgreater{}        & None or ``sequence''                  \\
15 & Data Type, Optionality & \textless{}D\_type\textgreater{}, optional                             & (int,optional)                      \\
16 & Structure, Optionality &\textless{}Structure\textgreater{}, optional                           & (array-like,optional)               \\
17 & Data Type, Values &\textless{}D\_type\textgreater \, or \textless{}value\textgreater{}      & int or ``all''                        \\
18 & Data Type, Values, Default Value &
  \textless{}D\_type\textgreater \, or \textless{}value\textgreater default=\textless{}value\textgreater{} & 
  int or ``all'', default=10 \\
  \hline
\end{tabular}
\end{table*}

%% file: figure/headerline.tex
\begin{figure}[t!]
\centering
  \includegraphics[width=\columnwidth]{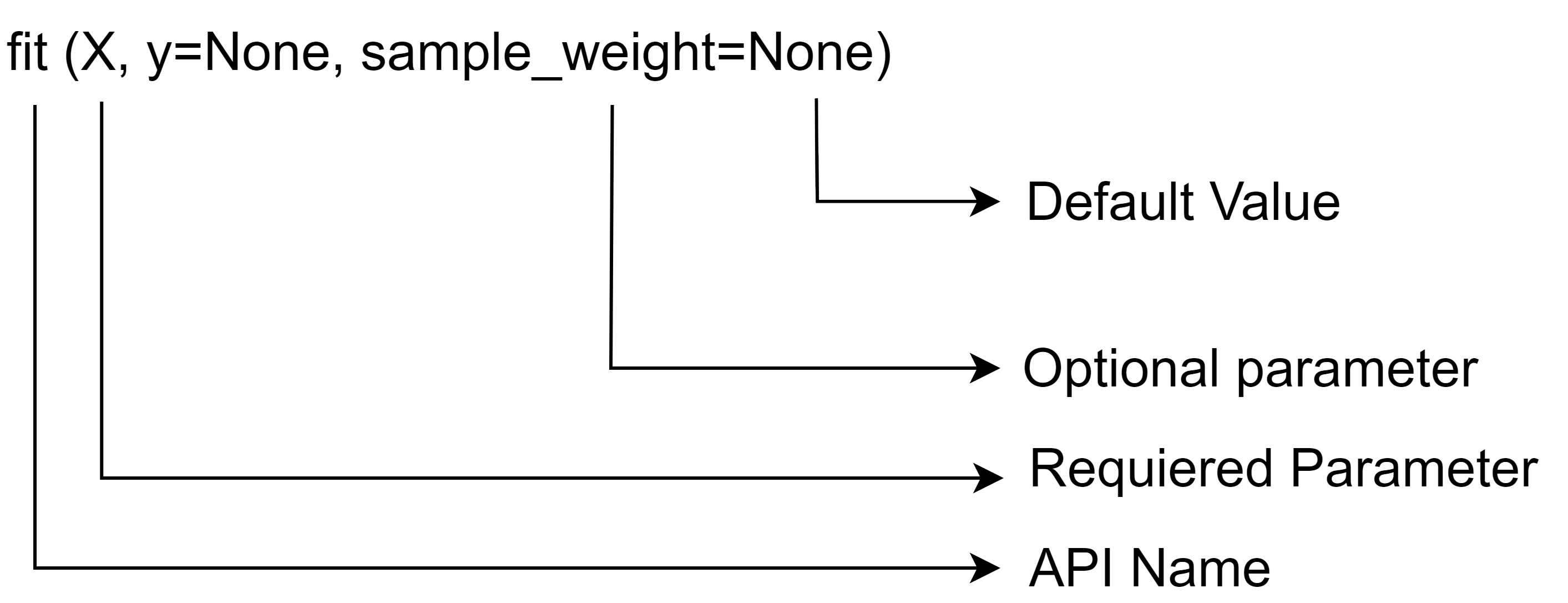}
  \caption{Signature definition of method \texttt{fit} from \texttt{sklearn.cluster.KMeans}}
 \label{fig:headerLine}
\end{figure}

%% file: figure/parameterLine.tex
\begin{figure}[t!]
\centering
  \includegraphics[width=\columnwidth]{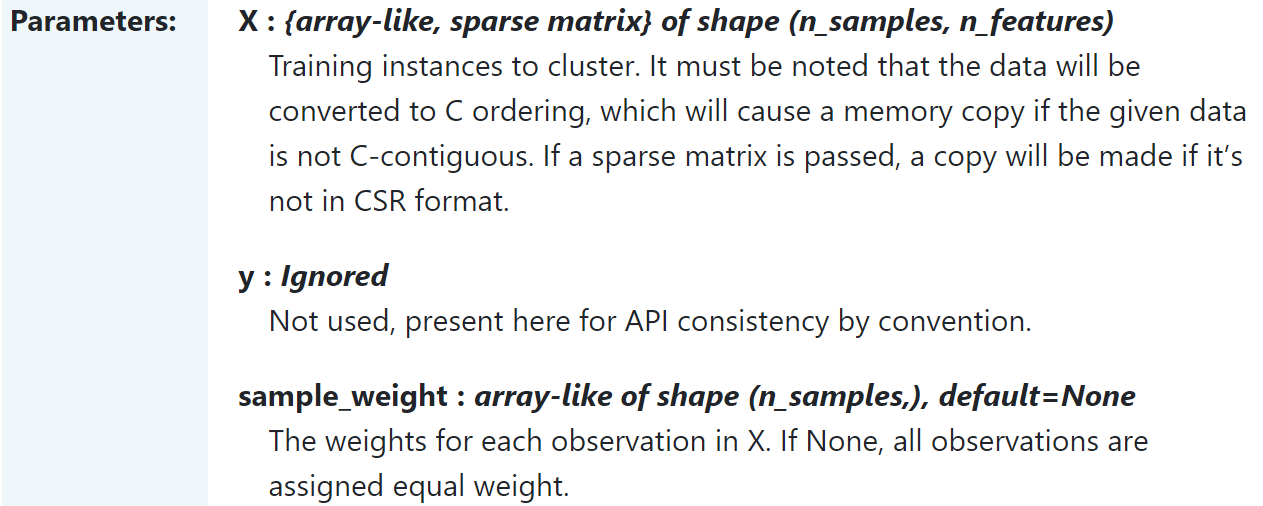}
  \caption{Parametric page of method \texttt{fit} from \texttt{sklearn.cluster.KMeans}}
  \label{fig:parametricPage}
\end{figure}

%% file: figure/codeExample.tex
\begin{figure}[t!]

	\begin{minipage}{\columnwidth}
    \begin{lstlisting}[language=Python]
from sklearn.cluster import KMeans
import numpy as np
X = np.array([[1, 2], [1, 4], [1, 0],
    [10, 2], [10, 4], [10, 0]])
kmeans = KMeans(n_clusters=2,
random_state=0).fit(X)
kmeans.predict([[0, 0], [12, 3]]) 
    \end{lstlisting}
    \vspace{-0.1in}
    \caption{Example code of \texttt{sklearn.cluster.KMeans}.}
    \label{figure:codeExamp}
    \end{minipage}
\end{figure}

%% file: table/constraintsFit.tex
\begin{table}[t!] 
\caption{API constraints extracted for API \texttt{fit} from \texttt{sklearn.cluster.KMeans} (Para is the parameter, Dvalue is the default value. Opt indicates whether the parameter is optional).}
\label{tab:constraintsExample}
\centering
\scalebox{0.85}{
\begin{tabular}{|c|c|c|c|c|c|c|}
\hline
\textbf{Para} & \textbf{Structure} & \textbf{Data  Type} & \textbf{Dvalue} & \textbf{Size} & \textbf{Shape} & \textbf{Opt} \\ \hline
X  & Array & Integer    & N/A & 2D    & (n,n)     & T    \\ \hline
y  & Undefined  & Undefined  & None     & N/A   & N/A   & F    \\ \hline
sample\_weight     & Array & float   & None     & N/A   & n     & F    \\ \hline
\end{tabular}
}
\end{table}

%% file: sec/setup.tex
\section{Experiment Setup}
\label{sec:experiment}
\input{table/projects}
\subsection{Research Questions}
\label{sec:rq}

{We answer the following research questions to evaluate the performance of our proposed approaches.}


\noindent {RQ1 (\textbf{Accuracy}): How accurate is the mined API knowledge from API documentations?}
\vspace{4pt}

\noindent {RQ2 (\textbf{Coverage Improvement}): Can {\tool} improve existing test generators for DL frameworks regarding code coverage?}
\vspace{4pt}

\noindent {RQ3 (\textbf{Invalid Test Removal}): Can {\tool} help remove the invalid tests generated by existing test generators for DL frameworks?}
\vspace{4pt}

\noindent {RQ4 (\textbf{Usefulness}): Are the generated test cases useful for developers?} 
\vspace{4pt}

In RQ1, we evaluate the accuracy of API constraints mined from the API documents. Note that, as the API usage patterns are mined from the accepted answers in SO, we assume they are correct and reflect different usage scenarios.  
In RQ2 and RQ3, we set out to whether the mined API knowledge can help improve test generation for DL frameworks regarding code coverage and invalid test case removals.  
In RQ4, we conduct a user case study to explore the practice value of {\tool} in generating test cases for DL frameworks. 

\subsection{Experiment Data}
\label{sec:dataset}


In this paper, we use API documentation and modules from four commonly used DL frameworks, i.e., Scikit-learn, PyTorch, TensorFlow, and CNTK throughout our analysis. 
Scikit-Learn is a typical machine learning and deep learning framework that provide APIs for various tasks. 
PyTorch provides high-level APIs to hide the low-level details of implementing deep learning applications. 
TensorFlow and CNTK were developed by Google and Microsoft respectively, are two of the most commonly used deep learning frameworks in the industry.  
For each of the DL frameworks, we extracted constraints of its APIs which have at least one parameter. 
Table~\ref{tab:projects} shows the details of each studied DL framework regarding the number of APIs, the total number of parameters of all the APIs, the number of SO posts, and the number of code examples in the API documents of the DL framework. 



\subsection{Evaluation Measure} 
\label{sec:metric}
Following existing studies~\cite{fraser2011evosuite,fraser2015does}, we utilize code coverage to evaluate the effectiveness of {\tool} in test case generation. 
Code coverage of test cases determines the percentage of the code under the test has been executed and tested. Code coverage can be at different levels, e.g., branch level, statement level, and method level. 
In this work, following the existing test generators, i.e., \textbf{PyEvosuite} and  \textbf{PyRandoop}~\cite{lukasczyk2020automated}, we measure the branch level code coverage with \texttt{coverage.py}\footnote{\url{https://coverage.readthedocs.io/en/6.3.2}}. We run both test generators and {\tool} on a 2.90GHz  i7-10700F desktop with 16GB of memory. 
Note that, {since our proposed {\tool} is designed as integrating the API knowledge with existing test generators, it can potentially help reduce the invalid tests generated by existing test generators}, we also examine to what extend can {\tool} help reduce the invalid tests generated by the two baseline approaches. 

%% file: table/projects.tex
\begin{table}[t]
\caption{Experiment DL frameworks used in this paper. \textbf{Ver} denotes the version, \textbf{\#Para} is the number of parameters.}
\centering
\label{tab:projects}
\begin{tabular}{l|l|c|c|c|c}
\hline
\textbf{Framework} & \textbf{Ver}  & \textbf{\#APIs} &\textbf{\#Para}&\textbf{\#Posts}&\#\textbf{Examples}\\ \hline
Sickit-Learn & {0.24.2}         &       1,210 & 4,980& 8,735 & 346 \\ \hline
PyTorch & 1.10.1 & 285 & 833& 4,161& 174\\ \hline
TensorFlow & 2.9.0 & 359 & 926& 17068 & 121\\ \hline
CNTK & 2.7 & 117 & 572& 397 & 92 \\ \hline
\end{tabular}
\end{table}

%% file: sec/result.tex
\section{Results and Analysis}
\label{sec:result}
This section presents the experimental results and answers the research questions in Section~\ref{sec:rq}.

\subsection{RQ1: Accuracy of Mined API Knowledge}
\label{sec:answer_rq2}


\textbf{Approach:} 
To answer this question, we first follow our API knowledge mining approaches proposed in (Section~\ref{sec:apiCons}) and Section~\ref{sec:mineAPIusage} to mine API constraints (i.e., structure, data type, default values, and shape/size of the structure) and API usage patterns from API documents and SO posts respectively for all the APIs in the studied four DL frameworks.  The detailed statistics of mined constraints are shown in Table~\ref{rq2constraints}. Overall, we extract more than 1.6K, 4.4k, 5.0k, 1.2k, and 1.1k constraints for data structure, data type, default values, the shape of data structure, and size of data structure respectively. 
The number of API usage patterns mined for different DL frameworks varies, e.g., our approach mines 1.2k API usage patterns for TensorFlow while only 138 patterns are mined for CNTK. This is because TensorFlow is a very popular DL framework and there are more questions in SO than in any other three DL frameworks, thus it has more code fragments and more API usage patterns are inferred.

To evaluate the accuracy of these API constraints mined, we 
manually check the signature definitions, parametric pages, and code examples to collect the ground-truth API constraints. After that authors work together to manually verify the mined constraints against the ground truth in the API documents. Note that as we mine the API usage patterns from the code fragments of accepted answers of SO questions, we assume all the API usage patterns are correct.


\input{table/num_constrainsts}
\input{table/rq2}
\textbf{Result:}  
Table~\ref{tab:rq2} shows the accuracy of mined constrains from each DL framework. Overall, the accuracy of constraint extraction is higher that 81\% across different frameworks. While the performance on different framework vary, e.g., the accuracy of `Shape' related constraint extracted from Scikit-learn is 100\% while on CNTK, the accuracy is only 78.15\%, this could be caused by the diversity of API document formats and also the quality of documents provided by different DL frameworks, for example most of CNTK's parametric pages do not mention the default values even if the corresponding parameters have default value settings. 

\mybox{The constraints mined from the studied DL frameworks have high accuracy, which shows the usefulness of our constraint extraction approaches {to facilitate the follow-up guidance for test generation}.}

\input{table/RQ1_table}

\subsection{RQ2: Improvement on Code Coverage}
\label{sec:answer_rq2}
\textbf{Approach:} 
To answer this RQ, we first run the existing test generators, i.e., \textbf{PyEvosuite} and \textbf{PyRandoop}, on the four studied DL frameworks. To generate tests for each module from the DL frameworks, we used the default configurations and set the time limit to 5 minutes as suggested by~\cite{lukasczyk2020automated,song2021how}. With the generated test cases, we further run \texttt{coverage.py} to collect the branch coverage on each module of each DL framework. Note that, as both \textbf{PyEvosuite} and \textbf{PyRandoop} can generate flaky tests or tests with syntax errors, we have removed all these tests following~\cite{lukasczyk2020automated, song2021how}, i.e., we first run each test case and  tests that throw \textit{SyntaxError} will be filtered out, we then execute each non-syntax test five times, and removed flaky tests from the executions. 
This process was repeated until all remaining tests passed five times. 
We then collect all the modules from each DL framework, we then input them, the API constraints mined from the modules, and the API usage patterns mined from SO into {\tool}, for each module we run {\tool} for five minutes and remove the generated flaky tests or tests with syntax errors as we did for \textbf{PyEvosuite} and \textbf{PyRandoop} to make the comparison fair.  
Note that as the test generation algorithms of \textbf{PyEvosuite}, \textbf{PyRandoop}, and our {\tool} contain randomness, which can affect their performance, we run the experiments of each tool 20 times and use the average code coverage to represent its performance. 


\input{table/RQ44}

\textbf{Result:}  
Table~\ref{rq1result} shows the results of code coverage analysis on test cases generated by \textbf{PyEvosuite}, \textbf{PyRandoop}, and  {\tool} on the modules across the four DL frameworks. For each DL framework, the table shows the number of generated tests and average code coverage achieved by the test generator. The average code coverage of a DL framework is calculated by calculating the arithmetic mean of coverage achieved by test cases generated for each API from it. The number of generated tests varies dramatically for different DL frameworks, this is mainly caused by the different API set sizes in these DL frameworks. 
Note that, in this experiment we only show the performance of {\tool} based on \textbf{PyEvosuite}, as \textbf{PyEvosuite} has better performance than \textbf{Randoop}. We have observed much improvement when build {\tool} on \textbf{Randoop}. 
On average, \textbf{PyEvosuite} achieves the coverage of 34.16\% on the four DL frameworks while the average coverage of \textbf{PyRandoop} is 30.87\%.  
Overall, {\tool} achieves better code coverage on each DL framework compared to \textbf{PyEvosuite}, the improvement ranges from 15.72\% (CNTK) to 27.0\% (Scikit Learn). 
Our Wilcoxon signed-rank
test ($p < 0.05$) also suggests that  {\tool} can significantly outperform both \textbf{PyEvosuite} and \textbf{Randoop}. 

We further combine the tests generated by {\tool} with the original tests of each DL framework to check whether these tests generated by {\tool} can help improve the code coverage. Table~\ref{tab:unittest} projects the code coverage improvements. 
Overall, the combined test set outperforms the code coverage by both the unit test cases and the test cases by {\tool}. The improvement ranges from 1.18\% to 7.05\%. Note that the performance of the unite test cases in the DL frameworks depends on several factors including the developer skills and maintenance of the software. So the coverage of these unit test case significantly differ from each DL framework, whereas the performance of the test cases developed by {\tool} depends on the availability of API knowledge. Our Wilcoxon signed-rank
test ($p < 0.05$) further suggests the significance of the improvement.

\vspace{4pt}
\mybox{{\tool} can significantly improve performance of test generation for DL frameworks. The improvement against existing test generators can be up to 27.0\% regarding code coverage. In addition, tests generated by {\tool} can further improve the code coverage of DL frameworks' original tests.}

\subsection{RQ3: Invalid Test Case Removal}

We have noticed that both \textbf{PyEvosuite} and \textbf{PyRandoop} generate many invalid tests. 
These invalid tests are neither flaky nor have syntax errors, yet they take incorrect inputs which violate the parameter constraints and {would not work as expected}.
For example, if an API with parameter $p$ only requires an integer variable, but the test case passes a string variable, then we call it an invalid test. As Python is a dynamic programming language, such tests could pass without throwing any errors, {yet could not produce the anticipated results.} {Since our proposed {\tool} is designed as integrating the API knowledge with existing test generators, it can potentially help reduce the invalid tests.} In this RQ, we explore whether {\tool} can help reduce the invalid tests generated by the two baseline approaches.  


\textbf{Approach:} We first randomly collect 100 tests generated by \textbf{PyEvosuite} on these four DL frameworks as \textbf{PyEvosuite} performs better than \textbf{PyRandoop}, then we manually check whether the generated tests are invalid. 
After that, for the verified invalid tests, we regenerate them with {\tool} and manually check whether {\tool} can correct their parameter issues.  

\textbf{Result:} 
Table~\ref{tab:invalidTests} shows the number of invalid tests generated by \textbf{PyEvosuite} and the number of invalid tests that {\tool} can help avoid. 
Overall, \textbf{PyEvosuite} generates around 55\% invalid tests across the four DL frameworks. We can also observe that the number of invalid tests on TensorFlow and Pytorch is larger than these of others. 
The possible reason is that both of provide APIs that require more complex inputs. We can also see that with the help of {\tool}, in total 42  out of the 221 invalid tests can be removed.  {\tool} fails to correct the left invalid tests mainly because of a lack of detailed data type information which is mined from code examples, as for most of the APIs there is no example code provided in the DL frameworks. {In future work, we plan to mine the code examples from other sources, e.g., GitHub, to retrieve more useful knowledge for the parameter settings.}

\mybox{{\tool} can help remove 19.0\% (42 out of 221) invalid tests generated by \textbf{PyEvosuite} on the four DL frameworks, which suggests the practice value of {\tool}.}

\input{table/RQ4}

\subsection{RQ4: Usefulness of the Generated Tests}
\label{sec:answer_rq5}

\noindent \textbf{Approach:} We further conduct a user study to investigate the usefulness of tests generated by {\tool} compared to tests generated by the existing test generators. Specifically, we first randomly selected five modules from each of the four DL frameworks and use both {\tool} and \textbf{PyEvosuite} to generate tests given 5 minutes. Note here, we only compare {\tool} to \textbf{PyEvosuite} as \textbf{PyEvosuite} outperforms \textbf{PyRandoop} on all the four examined projects as shown in  Section~\ref{sec:answer_rq2}. We also remove all the tests having syntax errors or being flaky. 
After that, we randomly select 10 tests from each module as our experiment subjects. We invited 16 students (i.e., six Ph.D. students and ten MS students) who are familiar with machine learning development and with at least 2 years of experience for our user study. We shuffled the test cases without showing the generator information during our experiment. For each test case, we ask the participators to score them from two aspects, i.e., \textit{Readability} and \textit{Usefulness}. 
\textit{Readability} measures to what extent the generated test case is easy to follow and easy to be understood. 
\textit{Usefulness} measures to what extent the generated test case is useful for developers to write unit tests for the given DL framework. 
The score ranges from 1 to 5. Score 1 means ``unreadable/useless'' and 5 means highly ``readable/useful''.

\input{figure/RQ4_boxplot}

\noindent \textbf{Result:} Figure~\ref{fig:casestudy} shows the boxplot results of our user case study. As we can see, tests generated by {\tool} have significantly better readability and usefulness than that of tests generated by \textbf{PyEvosuite}. Overall, the average scores of readability and usefulness for tests generated by {\tool} and \textbf{PyEvosuite} are 3.8 and 4.2, versus 2.7 and 2.1 respectively. 
We further used Wilcoxon signed-rank test for verifying the statistical significance of the differences. The p-values for both readability and usefulness time are small than 0.05, which suggests that {\tool} can generate significantly better test cases than \textbf{PyEvosuite} for DL frameworks.
\mybox{{\tool} can generate more readable and useful tests for the examined four DL frameworks, compared to the existing test generators.}


%% file: table/num_constrainsts.tex
\begin{table}[t!]
\centering
\caption{Statistics of API usage patterns and API constraints extracted for each DL framework. \textbf{\#Patterns} is number of API usage patterns mined from SO.}
 \label{rq2constraints}
\resizebox{0.98\columnwidth}{!}{
\begin{tabular}{|c|c|ccccc|}
\hline
\multirow{2}{*}{\textbf{Framework}} & \multirow{2}{*}{\textbf{\#Patterns}} & \multicolumn{5}{c|}{\textbf{\#API Constraints}} \\ \cline{3-7} 
 &  & \multicolumn{1}{c|}{\textbf{Structure}} & \multicolumn{1}{c|}{\textbf{Datatype}} & \multicolumn{1}{c|}{\textbf{Default}} & \multicolumn{1}{c|}{\textbf{Shape}} & \textbf{Size} \\ \hline
\textbf{Scikit Learn} & 341 & \multicolumn{1}{c|}{713} & \multicolumn{1}{c|}{3,490} & \multicolumn{1}{c|}{3,167} & \multicolumn{1}{c|}{649} & 504 \\ \hline
\textbf{Pytorch} & 172 & \multicolumn{1}{c|}{317} & \multicolumn{1}{c|}{371} & \multicolumn{1}{c|}{782} & \multicolumn{1}{c|}{214} & 206 \\ \hline 
\textbf{TensorFlow} & 1274 & \multicolumn{1}{c|}{351} & \multicolumn{1}{c|}{295} & \multicolumn{1}{c|}{696} & \multicolumn{1}{c|}{247} & 280 \\ \hline 
\textbf{CNTK} & 138 & \multicolumn{1}{c|}{227} & \multicolumn{1}{c|}{307} & \multicolumn{1}{c|}{417} & \multicolumn{1}{c|}{164} & 185 \\ \hline \hline
\textbf{Overall} & 1,925 & \multicolumn{1}{c|}{1,608} & \multicolumn{1}{c|}{4,463} & \multicolumn{1}{c|}{5,062} & \multicolumn{1}{c|}{1,274} &   1,175 \\ \hline
\end{tabular}%
}
\end{table}

%% file: table/rq2.tex
\begin{table}[t!]
\caption{The accuracy of mined constraints.}
\label{tab:rq2}
\resizebox{\columnwidth}{!}{%
\begin{tabular}{|l|lllll|}
\hline
\multirow{2}{*}{\textbf{Framework}} & \multicolumn{5}{c|}{\textbf{Accuracy of Constraints}}                                                              \\ \cline{2-6} 
                         &                     \multicolumn{1}{l|}{\textbf{Structure}} & \multicolumn{1}{l|}{\textbf{Data Type}} & \multicolumn{1}{l|}{\textbf{Shape}} & \multicolumn{1}{l|}{\textbf{Size}} & \textbf{Default} \\ \hline
                  Scikit-Learn          & \multicolumn{1}{l|}{91.72\% }      &\multicolumn{1}{l|}{95.28\% }      & \multicolumn{1}{l|}{100\%}      & \multicolumn{1}{l|}{100\% }      &90.75\%         \\ \hline
                              Pytorch       & \multicolumn{1}{l|}{90.33\% }      &\multicolumn{1}{l|}{92.75\% }      & \multicolumn{1}{l|}{87.14\%}      & \multicolumn{1}{l|}{87.14\% }      &97.27\%         \\ \hline
 TensorFlow   & \multicolumn{1}{l|}{81.47\%}      &\multicolumn{1}{l|}{ 76.20\%}      & \multicolumn{1}{l|}{80.13\%}      & \multicolumn{1}{l|}{84.50\% }      &  92.85\%       \\ \hline
  CNTK    & \multicolumn{1}{l|}{81.25\% }      &\multicolumn{1}{l|}{87.40\%}      & \multicolumn{1}{l|}{78.15\%}      & \multicolumn{1}{l|}{71.90\%}      &93.33\%       \\ \hline
\end{tabular}}
\end{table}

%% file: table/RQ1_table.tex
\begin{table*}[t!]
\centering
\caption{Performance of \textbf{PyRandoop}, \textbf{PyEvosuite}, and {\tool} on the studied DL frameworks.}
\label{rq1result}
\begin{tabular}{|l|cc|cc|c|c|c|}
\hline
\multirow{2}{*}{\textbf{Framework}} & \multicolumn{2}{c|}{\textbf{PyRandoop}}                   & \multicolumn{2}{c|}{\textbf{PyEvosuite}} 
& \multicolumn{3}{c|}{\textbf{{\tool}}} \\ \cline{2-8} 
 & \multicolumn{1}{l|}{\textbf{\#Test}} & \textbf{Coverage} & \multicolumn{1}{l|}{\textbf{\#Test}} & \textbf{Coverage} & \multicolumn{1}{l|}{\textbf{\#Test}} & \textbf{Coverage} & \textbf{Improvement}\\ \hline
Scikit Learn & \multicolumn{1}{l|}{1,573} &  43.52\%  & \multicolumn{1}{l|}{1,536}                &  49.28\% & \multicolumn{1}{l|}{1,588} & 62.59\% & 27.00\%\\ \hline
Pytorch  & \multicolumn{1}{l|}{492}                &      25.71\%             & \multicolumn{1}{l|}{558}                & 27.80\%  & \multicolumn{1}{l|}{529}   & 32.97\% & 18.60\%    \\ \hline

TensorFlow  & \multicolumn{1}{l|}{649}                &    23.18\%             & \multicolumn{1}{l|}{627}                & 24.07\%    & \multicolumn{1}{l|}{661} & 29.12\% &20.98\%             \\ \hline
CNTK  & \multicolumn{1}{l|}{257}                &    31.07\%             & \multicolumn{1}{l|}{281}                & 35.49\%       & \multicolumn{1}{l|}{318}   & 41.07\% & 15.72\%       \\
\hline \hline
\textbf{Average} & \multicolumn{1}{l|}{745.7}                &      30.87\%             & \multicolumn{1}{l|}{759.3}                & 34.16\% & 774.0 &41.44\% & 20.5\% \\\hline 
\end{tabular}
\end{table*}

%% file: table/RQ44.tex
\begin{table}[t!]

\caption{Code Coverage improvement with tests generated by {\tool} compared to the original tests.}
\label{tab:unittest}
\centering
\resizebox{0.98\columnwidth}{!}{%
\begin{tabular}{|c|c|cc|}
\hline
\multirow{2}{*}{\textbf{Framework}} & \textbf{Original tests} & \multicolumn{2}{c|}{\textbf{{\tool}$+$Original tests}} \\ \cline{2-4} 
 & \textbf{Coverage} & \multicolumn{1}{c|}{\textbf{Coverage}} & \textbf{Improvement} \\ \hline
Scikit Learn & 67.25\% & \multicolumn{1}{c|}{69.97\%} & 4.04\% \\ \hline
Pytorch & 58.47\% & \multicolumn{1}{c|}{59.16\%} & 1.18\% \\ \hline
Tensorflow & 43.65\% & \multicolumn{1}{c|}{46.73\%} & 7.05\% \\ \hline
CNTK & 54.29\% & \multicolumn{1}{c|}{56.34\%} & 3.77\% \\ \hline 
\end{tabular}%
}
\end{table}

%% file: table/RQ4.tex
\begin{table}[t!]
\caption{Invalid tests generated by \textbf{PyEvosuite} and can be removed by {\tool}.}
\label{tab:invalidTests}
\centering
\begin{tabular}{|l|c|c|}
\hline
\textbf{Framework} & \textbf{Invalid Tests} & \textbf{Removed by {\tool}} \\ \hline
Scikit Learn & 43 &  9   \\ \hline
Pytorch & 59 &  14   \\ \hline 
TensorFlow & 68 & 7    \\ \hline
CNTK & 51 &  12   \\ \hline
\hline
Overall & 221 & 42 \\ \hline
\end{tabular}
\end{table}

%% file: figure/RQ4_boxplot.tex
\begin{figure}[t!]
\begin{subfigure}{0.48\linewidth}
\includegraphics[width=\linewidth]{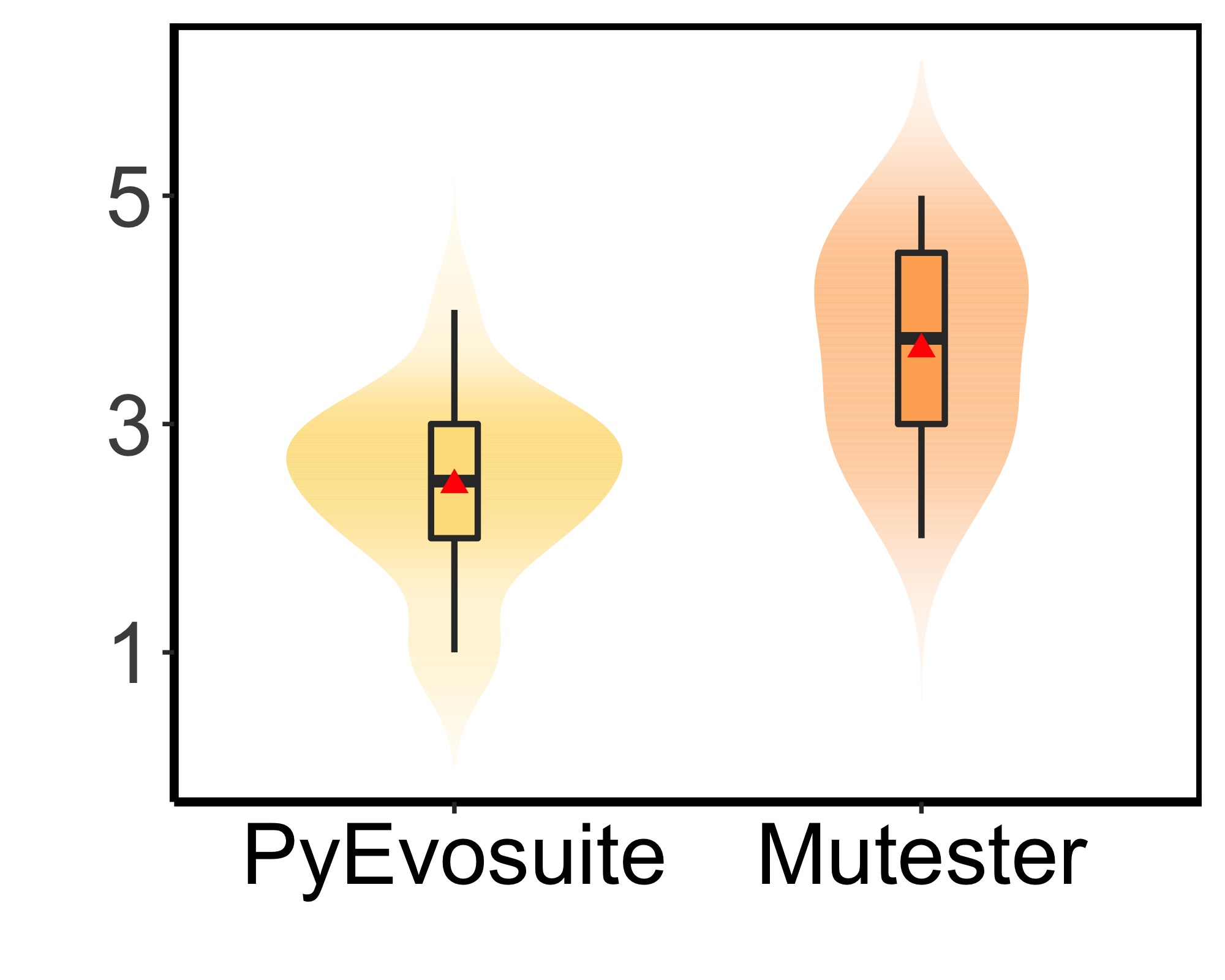}
\caption{\small Readability}
\end{subfigure} 
\begin{subfigure}{0.48\linewidth}
\centering
\includegraphics[width=\linewidth]{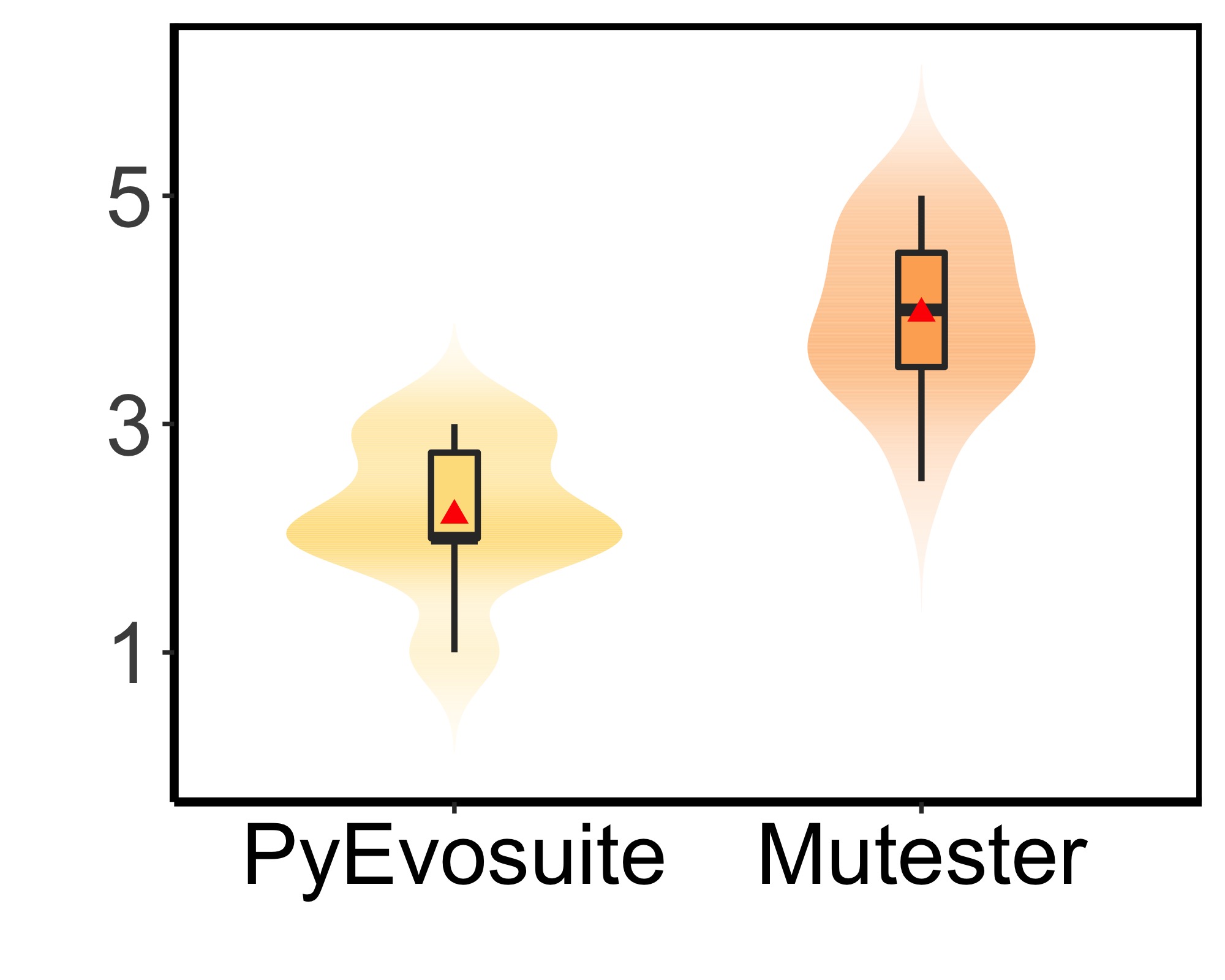}
\caption{\small Usefulness}
\end{subfigure}
\caption{Results of user study.}
\label{fig:casestudy}
\end{figure}

%% file: sec/discussion.tex
\section{Threats to Validity}
\label{sec:discussion}




In this paper, we use Evosuite and Randoop algorithms implemented by the authors of Evosuite in~\cite{lukasczyk2020automated} as our experiment subjects, so the correctness of their implementations can affect the performance of {\tool}. We may get different results using different unit test case generation tools. Nonetheless, these tools are widely used and accepted ones both in academia and industry, and their generated tests are similar to the modern test case generation tools including TestFul~\cite{baresi2010testful}, JavaPathFinder~\cite{visser2004test}, Pex~\cite{tillmann2008pex}, and others. 
We use code coverage to assess the efficiency of generated test cases based on the assumption that code coverage and test effectiveness has a correlation with each other. There are other measures for this purpose such as mutation score, which we plan to use to examine the effectiveness of tests in the future. 
The correctness of constraints acquired from mining API documents is based on the correctness of the document itself. If there is some wrong information about the API in the documents, our tool does not identify them. It cannot be guaranteed that all the code snippets mined from Stack Overflow correctly represent the usage of an or set of APIs, as they are constructed by programmers of various levels of expertise. 
In this paper, we have conducted all experiments based on open source deep learning frameworks written in Python language. These DL frameworks are commonly used and widely accepted across many application domains. Our findings in this paper may not be generalizable to commercial DL frameworks. We plan to extend the proposed test case generation tool for other programming languages in the future.

%% file: sec/related.tex
\section{Related Work}
\label{sec:related}

\subsection{Automatic Test Case Generation}
\label{sec:testGeneration}

To help developers generate tests, many automatic approaches have been proposed~\cite{derakhshanfar2020search,olsthoorn2020generating,mcminn2012search,shahbaz2015automatic,arcuri2021evomaster,atlidakis2019restler,pacheco2005eclat,fraser2011evosuite,pacheco2007randoop,csallner2004jcrasher,pond2005hyphy,shaffer1995multiple}. 
JCrasher~\cite{csallner2004jcrasher} is one of the first random based test generator, which creates Java-based method call sequences that throw certain exceptions. Similarly Ecalt~\cite{pacheco2005eclat} and Randoop~\cite{pacheco2007randoop} employed the idea of random search to generate tests that can expose more faults. The major limitation of random testing is that it has no guidance about testing which results in low code coverage. 

To augment random testing, numerous approaches have been proposed including search-based algorithms and symbolic execution~\cite{fraser2011evosuite,baresi2010testful,leitner2007contract}. eToc~\cite{tonella2004evolutionary} first employed genetic algorithms to test primitive data types and strings by generating test data. Along this line, Fraser et al.~\cite{fraser2011evosuite} proposed EvoSuite, which uses a genetic algorithm for automatic test case generation. The objective function is maximizing code coverage based on a set of pre-defined criteria, e.g. covering all branches or statements, using evolutionary algorithms. Symbolic execution based test generation~\cite{56,57} is another approach to extending random testing where a program is executed abstractly in such a way that all possible combinations of inputs to a program is covered by considering execution path in a source code. One of the typical symbolic execution based test generation tools is Pex~\cite{tillmann2008pex}, it can automatically produce parameterized unit tests with high code coverage for a .NET program via a systematic program analysis to determine test inputs. Pex learnt the program behavior by monitoring execution traces. However, symbolic execution based approaches often suffer from a lack of scalability. 


Recently fuzz testing based tools have been proposed to find bugs of DL frameworks~\cite{guo2020audee,xie2022docter,deng2022fuzzing,kang2022skipfuzz}. These tools and our approach have different targets and propose different techniques. Fuzz testing tools use different techniques to generate valid and invalid inputs for a given API of DL frameworks and then run the API with the generated inputs with the purpose of crashing the API for bug detection. 
Our approach mainly focuses on generating unit test cases for APIs of DL frameworks by leveraging the API constraints mined from API documentations and API usage patterns mined from SO.

\subsection{API Knowledge Mining}
\label{sec:apiDoc}

API documents and Stack Overflow are two important sources for API knowledge mining~\cite{treude2016augmenting}. 
There are many existing studies that analyze API usage scenarios and extract useful patterns from API documents for specific tasks, e.g., API usage mining~\cite{zhong2017icse,zhong2009mapo,maalej2013patterns,liu2021learning}, API misuse detection~\cite{nielebock2021guided,sven2019investigating}, 
API constraints mining for fuzzing testing~\cite{xie2011testing,li2020documentation},cloud API testing\cite{8029894}. For example, Zhang et al.~\cite{zhong2009mapo} developed an API usage mining framework equipped with a tool called MAPO (Mining API usage Pattern from Open source repositories) for automatic mining of API usage patterns. Based on programs' requests, MAPO recommends the mined API usage patterns and their associated code snippets. The experimental results indicate that MAPO is effective at assisting programmers in programming tasks. Another interesting work is conducted by 
Nielebock et al.~\cite{nielebock2021guided} mined API usage patterns to detect misuses in JAVA API used in open source projects. 
Li et al.~\cite{li2018improving} used NLP methods to discover API caveat sentences from API documentation and map them to entities in the API knowledge graph. To evaluate the approach, the authors build an API caveat knowledge graph on top of API documentation from Android applications. 
Maalej and Robillard~\cite{maalej2013patterns} analyzed the subtle knowledge in API reference documentation to understand its structure and nature by mining many APIs existing in two well-known platforms namely JAVA and C\#. 
Xie et al.~\cite{xie2022docter} leveraged sequential pattern mining to generate rules for extracting deep learning specific constraints from API documents and uses these constraints to guide the fuzzing testing of deep learning frameworks. The main difference between this work and our study is two-fold, 1) we mine API knowledge from two different sources, i.e., API documents and Stack Overflow while they mine API constrains only from API documents. 2) we leverage the mined API knowledge to improve test case generation while they use the API constraints to guide fuzzing input generation.

The automated mining of crowd-sourced knowledge from SO has generated considerable attention in recent years~\cite{calefato2015mining,uddin2020mining,treude2016augmenting,huang2018api}. Uddin et al.~\cite{uddin2020mining} mined API documentation from stack overflow using automated techniques to analyse API usage scenarios. 
Treude et al.~\cite{treude2016augmenting} proposed an approach to augmenting API documentation with API knowledge mined from stack overflow. Huang et al.~\cite{huang2018api} mined the API usage from SO to help recommend APIs for new queries posted in SO. Different from these studies, in this work we mine the API usage of API from DL frameworks to help test generation.

%% file: sec/conclusion.tex
\section{Conclusion}
\label{sec:conclusion}
This paper proposes {\tool} to generate test cases for APIs of DL frameworks by leveraging the API constraints mined from the corresponding API documentations and the API usage patterns mined from code fragments in Stack Overflow (SO). 
Given an API, {\tool} combines its API knowledge with existing test generators to generate test cases to test the respective API. 
Results of our experiment on four widely-used DL frameworks (i.e., Scikit-learn, PyTorch, TensorFlow, and CNTK) show that {\tool} can significantly improve the corresponding test generation methods and the improvement in code coverage is 15.7\% to 27.0\% on average. In addition, it can also help reduce around 19.0\% of invalid tests generated by the existing test generators. 
{Our user study with 16 developers further demonstrates the practicality of {\tool} in generating unit test cases for DL frameworks.}

%% file: main.bbl
\begin{thebibliography}{10}
\providecommand{\url}[1]{#1}
\csname url@samestyle\endcsname
\providecommand{\newblock}{\relax}
\providecommand{\bibinfo}[2]{#2}
\providecommand{\BIBentrySTDinterwordspacing}{\spaceskip=0pt\relax}
\providecommand{\BIBentryALTinterwordstretchfactor}{4}
\providecommand{\BIBentryALTinterwordspacing}{\spaceskip=\fontdimen2\font plus
\BIBentryALTinterwordstretchfactor\fontdimen3\font minus
  \fontdimen4\font\relax}
\providecommand{\BIBforeignlanguage}[2]{{%
\expandafter\ifx\csname l@#1\endcsname\relax
\typeout{** WARNING: IEEEtran.bst: No hyphenation pattern has been}%
\typeout{** loaded for the language `#1'. Using the pattern for}%
\typeout{** the default language instead.}%
\else
\language=\csname l@#1\endcsname
\fi
#2}}
\providecommand{\BIBdecl}{\relax}
\BIBdecl

\bibitem{fraser2011evosuite}
G.~Fraser and A.~Arcuri, ``Evosuite: automatic test suite generation for
  object-oriented software,'' in \emph{FSE'11}, 2011, pp. 416--419.

\bibitem{pacheco2007randoop}
C.~Pacheco and M.~D. Ernst, ``Randoop: feedback-directed random testing for
  java,'' in \emph{Companion to the 22nd ACM SIGPLAN conference on
  Object-oriented programming systems and applications companion}, 2007, pp.
  815--816.

\bibitem{pacheco2007feedback}
C.~Pacheco, S.~K. Lahiri, M.~D. Ernst, and T.~Ball, ``Feedback-directed random
  test generation,'' in \emph{29th International Conference on Software
  Engineering (ICSE'07)}, 2007, pp. 75--84.

\bibitem{song2021how}
S.~Wang, N.~Shrestha, A.~K. Subburaman, J.~Wang, M.~Wei, and N.~Nagappan,
  ``Automatic unit test generation for machine learning libraries: How far are
  we?'' in \emph{2021 IEEE/ACM 43rd International Conference on Software
  Engineering (ICSE)}, 2021, pp. 1548--1560.

\bibitem{zhang2020machine}
J.~M. Zhang, M.~Harman, L.~Ma, and Y.~Liu, ``Machine learning testing: Survey,
  landscapes and horizons,'' \emph{IEEE Transactions on Software Engineering},
  2020.

\bibitem{lukasczyk2020automated}
S.~Lukasczyk, F.~Kroi{\ss}, and G.~Fraser, ``Automated unit test generation for
  python,'' in \emph{International Symposium on Search Based Software
  Engineering}.\hskip 1em plus 0.5em minus 0.4em\relax Springer, 2020, pp.
  9--24.

\bibitem{journals/corr/abs-2202-05218}
S.~Lukasczyk and G.~Fraser, ``Pynguin: Automated unit test generation for
  python,'' \emph{CoRR}, vol. abs/2202.05218, 2022.

\bibitem{sklearn_api}
L.~Buitinck, G.~Louppe, M.~Blondel, F.~Pedregosa, A.~Mueller, O.~Grisel,
  V.~Niculae, P.~Prettenhofer, A.~Gramfort, J.~Grobler, R.~Layton,
  J.~VanderPlas, A.~Joly, B.~Holt, and G.~Varoquaux, ``{API} design for machine
  learning software: experiences from the scikit-learn project,'' in \emph{ECML
  PKDD Workshop: Languages for Data Mining and Machine Learning}, 2013, pp.
  108--122.

\bibitem{NEURIPS2019_9015}
A.~Paszke, S.~Gross, F.~Massa, A.~Lerer, J.~Bradbury, G.~Chanan, T.~Killeen,
  Z.~Lin, N.~Gimelshein, L.~Antiga, A.~Desmaison, A.~Kopf, E.~Yang, Z.~DeVito,
  M.~Raison, A.~Tejani, S.~Chilamkurthy, B.~Steiner, L.~Fang, J.~Bai, and
  S.~Chintala, ``Pytorch: An imperative style, high-performance deep learning
  library.''

\bibitem{abadi2016tensorflow}
M.~Abadi, P.~Barham, J.~Chen, Z.~Chen, A.~Davis, J.~Dean, M.~Devin,
  S.~Ghemawat, G.~Irving, M.~Isard \emph{et~al.}, ``Tensorflow: A system for
  large-scale machine learning,'' in \emph{12th $\{$USENIX$\}$ Symposium on
  Operating Systems Design and Implementation ($\{$OSDI$\}$ 16)}, 2016, pp.
  265--283.

\bibitem{seide2016cntk}
F.~Seide and A.~Agarwal, ``Cntk: Microsoft's open-source deep-learning
  toolkit,'' in \emph{Proceedings of the 22nd ACM SIGKDD international
  conference on knowledge discovery and data mining}, 2016, pp. 2135--2135.

\bibitem{fraser2012whole}
G.~Fraser and A.~Arcuri, ``Whole test suite generation,'' \emph{IEEE
  Transactions on Software Engineering}, vol.~39, no.~2, pp. 276--291, 2012.

\bibitem{derakhshanfar2020search}
P.~Derakhshanfar, X.~Devroey, G.~Perrouin, A.~Zaidman, and A.~van Deursen,
  ``Search-based crash reproduction using behavioural model seeding,''
  \emph{Software Testing, Verification and Reliability}, vol.~30, no.~3, p.
  e1733, 2020.

\bibitem{olsthoorn2020generating}
M.~Olsthoorn, A.~van Deursen, and A.~Panichella, ``Generating highly-structured
  input data by combining search-based testing and grammar-based fuzzing,'' in
  \emph{2020 35th IEEE/ACM International Conference on Automated Software
  Engineering (ASE)}.\hskip 1em plus 0.5em minus 0.4em\relax IEEE, 2020, pp.
  1224--1228.

\bibitem{mcminn2012search}
P.~McMinn, M.~Shahbaz, and M.~Stevenson, ``Search-based test input generation
  for string data types using the results of web queries,'' in \emph{2012 IEEE
  Fifth International Conference on Software Testing, Verification and
  Validation}.\hskip 1em plus 0.5em minus 0.4em\relax IEEE, 2012, pp. 141--150.

\bibitem{shahbaz2015automatic}
M.~Shahbaz, P.~McMinn, and M.~Stevenson, ``Automatic generation of valid and
  invalid test data for string validation routines using web searches and
  regular expressions,'' \emph{Science of Computer Programming}, vol.~97, pp.
  405--425, 2015.

\bibitem{arcuri2021evomaster}
A.~Arcuri, J.~P. Galeotti, B.~Marculescu, and M.~Zhang, ``Evomaster: A
  search-based system test generation tool,'' \emph{Journal of Open Source
  Software}, vol.~6, no.~57, p. 2153, 2021.

\bibitem{atlidakis2019restler}
V.~Atlidakis, P.~Godefroid, and M.~Polishchuk, ``Restler: Stateful rest api
  fuzzing,'' in \emph{2019 IEEE/ACM 41st International Conference on Software
  Engineering (ICSE)}.\hskip 1em plus 0.5em minus 0.4em\relax IEEE, 2019, pp.
  748--758.

\bibitem{arcuri2011random}
A.~Arcuri, M.~Z. Iqbal, and L.~Briand, ``Random testing: Theoretical results
  and practical implications,'' \emph{TSE'11}, vol.~38, no.~2, pp. 258--277,
  2011.

\bibitem{9054844}
Y.~Hashemi, M.~Nayebi, and G.~Antoniol, ``Documentation of machine learning
  software,'' in \emph{2020 IEEE 27th International Conference on Software
  Analysis, Evolution and Reengineering (SANER)}, 2020, pp. 666--667.

\bibitem{richardson2007beautiful}
L.~Richardson, ``Beautiful soup documentation,'' \emph{April}, 2007.

\bibitem{chowdhary2020natural}
K.~Chowdhary, ``Natural language processing,'' \emph{Fundamentals of artificial
  intelligence}, pp. 603--649, 2020.

\bibitem{shamshiri2015automatically}
S.~Shamshiri, R.~Just, J.~M. Rojas, G.~Fraser, P.~McMinn, and A.~Arcuri, ``Do
  automatically generated unit tests find real faults? an empirical study of
  effectiveness and challenges (t),'' in \emph{ASE'15}, 2015, pp. 201--211.

\bibitem{zhang2018code}
T.~Zhang, G.~Upadhyaya, A.~Reinhardt, H.~Rajan, and M.~Kim, ``Are code examples
  on an online q\&a forum reliable?: a study of api misuse on stack overflow,''
  in \emph{2018 IEEE/ACM 40th International Conference on Software Engineering
  (ICSE)}.\hskip 1em plus 0.5em minus 0.4em\relax IEEE, 2018, pp. 886--896.

\bibitem{uddin2020mining}
G.~Uddin, F.~Khomh, and C.~K. Roy, ``Mining api usage scenarios from stack
  overflow,'' \emph{Information and Software Technology}, vol. 122, p. 106277,
  2020.

\bibitem{huang2018api}
Q.~Huang, X.~Xia, Z.~Xing, D.~Lo, and X.~Wang, ``Api method recommendation
  without worrying about the task-api knowledge gap,'' in \emph{2018 33rd
  IEEE/ACM International Conference on Automated Software Engineering
  (ASE)}.\hskip 1em plus 0.5em minus 0.4em\relax IEEE, 2018, pp. 293--304.

\bibitem{zhong2009mapo}
H.~Zhong, T.~Xie, L.~Zhang, J.~Pei, and H.~Mei, ``Mapo: Mining and recommending
  api usage patterns,'' in \emph{European Conference on Object-Oriented
  Programming}.\hskip 1em plus 0.5em minus 0.4em\relax Springer, 2009, pp.
  318--343.

\bibitem{agrawal1994fast}
R.~Agrawal, R.~Srikant \emph{et~al.}, ``Fast algorithms for mining association
  rules,'' in \emph{Proc. 20th int. conf. very large data bases, VLDB}, vol.
  1215.\hskip 1em plus 0.5em minus 0.4em\relax Citeseer, 1994, pp. 487--499.

\bibitem{fraser2015does}
G.~Fraser, M.~Staats, P.~McMinn, A.~Arcuri, and F.~Padberg, ``Does automated
  unit test generation really help software testers? a controlled empirical
  study,'' \emph{TOSEM'15}, vol.~24, no.~4, pp. 1--49, 2015.

\bibitem{baresi2010testful}
L.~Baresi, P.~L. Lanzi, and M.~Miraz, ``Testful: an evolutionary test approach
  for java,'' in \emph{ICST'10}, 2010, pp. 185--194.

\bibitem{visser2004test}
W.~Visser, C.~S. Pǎsǎreanu, and S.~Khurshid, ``Test input generation with
  java pathfinder,'' in \emph{Proceedings of the 2004 ACM SIGSOFT international
  symposium on Software testing and analysis}, 2004, pp. 97--107.

\bibitem{tillmann2008pex}
N.~Tillmann and J.~d. Halleux, ``Pex--white box test generation for. net,'' in
  \emph{International conference on tests and proofs}.\hskip 1em plus 0.5em
  minus 0.4em\relax Springer, 2008, pp. 134--153.

\bibitem{pacheco2005eclat}
C.~Pacheco and M.~D. Ernst, ``Eclat: Automatic generation and classification of
  test inputs,'' in \emph{European Conference on Object-Oriented
  Programming}.\hskip 1em plus 0.5em minus 0.4em\relax Springer, 2005, pp.
  504--527.

\bibitem{csallner2004jcrasher}
C.~Csallner and Y.~Smaragdakis, ``Jcrasher: an automatic robustness tester for
  java,'' \emph{Software: Practice and Experience}, vol.~34, no.~11, pp.
  1025--1050, 2004.

\bibitem{pond2005hyphy}
S.~L.~K. Pond and S.~V. Muse, ``Hyphy: hypothesis testing using phylogenies,''
  in \emph{Statistical methods in molecular evolution}.\hskip 1em plus 0.5em
  minus 0.4em\relax Springer, 2005, pp. 125--181.

\bibitem{shaffer1995multiple}
J.~P. Shaffer, ``Multiple hypothesis testing,'' \emph{Annual review of
  psychology}, vol.~46, no.~1, pp. 561--584, 1995.

\bibitem{leitner2007contract}
A.~Leitner, I.~Ciupa, M.~Oriol, B.~Meyer, and A.~Fiva, ``Contract driven
  development= test driven development-writing test cases,'' in
  \emph{Proceedings of the the 6th joint meeting of the European software
  engineering conference and the ACM SIGSOFT symposium on The foundations of
  software engineering}, 2007, pp. 425--434.

\bibitem{tonella2004evolutionary}
P.~Tonella, ``Evolutionary testing of classes,'' \emph{ACM SIGSOFT Software
  Engineering Notes}, vol.~29, no.~4, pp. 119--128, 2004.

\bibitem{56}
J.~C. King, ``Symbolic execution and program testing,'' \emph{Communications of
  the ACM}, vol.~19, no.~7, pp. 385--394, 1976.

\bibitem{57}
C.~S. P{\u{a}}s{\u{a}}reanu and N.~Rungta, ``Symbolic pathfinder: symbolic
  execution of java bytecode,'' in \emph{Proceedings of the IEEE/ACM
  international conference on Automated software engineering}, 2010, pp.
  179--180.

\bibitem{guo2020audee}
Q.~Guo, X.~Xie, Y.~Li, X.~Zhang, Y.~Liu, X.~Li, and C.~Shen, ``Audee: Automated
  testing for deep learning frameworks,'' in \emph{Proceedings of the 35th
  IEEE/ACM International Conference on Automated Software Engineering}, 2020,
  pp. 486--498.

\bibitem{xie2022docter}
D.~Xie, Y.~Li, M.~Kim, H.~V. Pham, L.~Tan, X.~Zhang, and M.~W. Godfrey,
  ``Docter: documentation-guided fuzzing for testing deep learning api
  functions,'' in \emph{Proceedings of the 31st ACM SIGSOFT International
  Symposium on Software Testing and Analysis}, 2022, pp. 176--188.

\bibitem{deng2022fuzzing}
Y.~Deng, C.~Yang, A.~Wei, and L.~Zhang, ``Fuzzing deep-learning libraries via
  automated relational api inference,'' in \emph{Proceedings of the 30th ACM
  Joint European Software Engineering Conference and Symposium on the
  Foundations of Software Engineering}, 2022, pp. 44--56.

\bibitem{kang2022skipfuzz}
H.~J. Kang, P.~Rattanukul, S.~A. Haryono, T.~G. Nguyen, C.~Ragkhitwetsagul,
  C.~Pasareanu, and D.~Lo, ``Skipfuzz: Active learning-based input selection
  for fuzzing deep learning libraries,'' \emph{arXiv preprint
  arXiv:2212.04038}, 2022.

\bibitem{treude2016augmenting}
C.~Treude and M.~P. Robillard, ``Augmenting api documentation with insights
  from stack overflow,'' in \emph{2016 IEEE/ACM 38th International Conference
  on Software Engineering (ICSE)}.\hskip 1em plus 0.5em minus 0.4em\relax IEEE,
  2016, pp. 392--403.

\bibitem{zhong2017icse}
H.~Zhong and N.~Meng, ``An empirical study on using hints from past fixes,'' in
  \emph{Proc. ICSE}, 2017, pp. 144--145.

\bibitem{maalej2013patterns}
W.~Maalej and M.~P. Robillard, ``Patterns of knowledge in api reference
  documentation,'' \emph{IEEE Transactions on Software Engineering}, vol.~39,
  no.~9, pp. 1264--1282, 2013.

\bibitem{liu2021learning}
M.~Liu, X.~Peng, A.~Marcus, C.~Treude, X.~Bai, G.~Lyu, J.~Xie, and X.~Zhang,
  ``Learning-based extraction of first-order logic representations of api
  directives,'' in \emph{Proceedings of the 29th ACM Joint Meeting on European
  Software Engineering Conference and Symposium on the Foundations of Software
  Engineering}, 2021, pp. 491--502.

\bibitem{nielebock2021guided}
S.~Nielebock, R.~Heum{\"u}ller, K.~M. Schott, and F.~Ortmeier, ``Guided pattern
  mining for api misuse detection by change-based code analysis,''
  \emph{Automated Software Engineering}, vol.~28, no.~2, pp. 1--48, 2021.

\bibitem{sven2019investigating}
A.~Sven, H.~A. Nguyen, S.~Nadi, T.~N. Nguyen, and M.~Mezini, ``Investigating
  next steps in static api-misuse detection,'' in \emph{2019 IEEE/ACM 16th
  International Conference on Mining Software Repositories (MSR)}.\hskip 1em
  plus 0.5em minus 0.4em\relax IEEE, 2019, pp. 265--275.

\bibitem{xie2011testing}
X.~Xie, J.~W. Ho, C.~Murphy, G.~Kaiser, B.~Xu, and T.~Y. Chen, ``Testing and
  validating machine learning classifiers by metamorphic testing,''
  \emph{JSS'11}, vol.~84, no.~4, pp. 544--558, 2011.

\bibitem{li2020documentation}
Y.~Li, ``Documentation-guided fuzzing for testing deep learning api
  functions,'' Master's thesis, University of Waterloo, 2020.

\bibitem{8029894}
J.~Wang, X.~Bai, L.~Li, Z.~Ji, and H.~Ma, ``A model-based framework for cloud
  api testing,'' in \emph{2017 IEEE 41st Annual Computer Software and
  Applications Conference (COMPSAC)}, vol.~2, 2017, pp. 60--65.

\bibitem{li2018improving}
H.~Li, S.~Li, J.~Sun, Z.~Xing, X.~Peng, M.~Liu, and X.~Zhao, ``Improving api
  caveats accessibility by mining api caveats knowledge graph,'' in \emph{2018
  IEEE International Conference on Software Maintenance and Evolution
  (ICSME)}.\hskip 1em plus 0.5em minus 0.4em\relax IEEE, 2018, pp. 183--193.

\bibitem{calefato2015mining}
F.~Calefato, F.~Lanubile, M.~C. Marasciulo, and N.~Novielli, ``Mining
  successful answers in stack overflow,'' in \emph{2015 IEEE/ACM 12th Working
  Conference on Mining Software Repositories}.\hskip 1em plus 0.5em minus
  0.4em\relax IEEE, 2015, pp. 430--433.

\end{thebibliography}
